\documentclass[nofootinbib,prd,showpacs]{revtex4}
\usepackage{amsmath}
\usepackage{subfigure}
\usepackage{amsfonts}
\usepackage{hyperref}
\usepackage{amssymb}
\usepackage{graphicx}
\usepackage[latin1]{inputenc}
\usepackage[english]{babel}
\usepackage{color,xcolor}

\definecolor{purple}{rgb}{1,0,1}
\definecolor{lime}{HTML}{A6CE39} % needs xcolor

\def\eqn#1{\eq\eqref{#1}}
\def\rf{\eqref}
\def\qq{\qquad}

\def\inch{\hspace*{1in}}

\def\lal{&&\ {}}
\def\eq{Eq.\,}
\def\eqs{Eqs.\,}
\def\beq{\begin{equation}}
\def\eeq{\end{equation}}
\def\bear{\begin{eqnarray}}
\def\bearr{\begin{eqnarray} \lal}
\def\ear{\end{eqnarray}}
\def\earn{\nonumber \end{eqnarray}}
\def\nn{\nonumber\\ {}}

\def\nnn{\nonumber\\ \lal }

\def\yyy{\\[5pt] \lal }

\def\const{{\rm const}}
\def\diag{\,{\rm diag}\,}

\def\d{\partial}
\def\mn{_{\mu\nu}}
\def\MN{^{\mu\nu}}
\def\mN{_\mu^\nu}

\def\R{{\mathbb R}}

\def\cK{{\mathcal K}}

\def\GR{general relativity}

\def\ssph{static, spherically symmetric}
\def\cy{cylindrically symmetric}
\def\scy{static, cylindrically symmetric}
\def\bh{black hole}
\def\bhs{black holes}
\def\wh{wormhole}

\def\emag{electromagnetic}
\def\Scw{Schwarz\-schild}
\def\RN{Reiss\-ner-Nord\-str\"om}
\begin{document}

\title{Cylindrical black bounces and their field sources}

%=================================================
\author{Kirill A. Bronnikov}
\email{kb20@yandex.ru}
\affiliation{Center of Gravitation and Fundamental Metrology, VNIIMS, Ozyornaya ulitsa 46, Moscow 119361, Russia}
\affiliation{Peoples' Friendship University of Russia, ulitsa Miklukho-Maklaya 6, Moscow, 117198, Russia}
\affiliation{National Research Nuclear University ``MEPhI'', Kashirskoe shosse 31, Moscow 115409, Russia}
%======================================================
\author{Manuel E. Rodrigues}
\email{esialg@gmail.com}
\affiliation{Faculdade de Ci\^{e}ncias Exatas e Tecnologia, 
Universidade Federal do Par\'{a}\\
Campus Universit\'{a}rio de Abaetetuba, 68440-000, Abaetetuba, Par\'{a}, Brazil}
\affiliation{Faculdade de F\'{\i}sica, Programa de P\'{o}s-Gradua\c{c}\~ao em 
F\'isica, Universidade Federal do 
 Par\'{a}, 66075-110, Bel\'{e}m, Par\'{a}, Brazil}
%======================================================
\author{Marcos V. de S. Silva}
\email{marco2s303@gmail.com}
\affiliation{Faculdade de F\'{\i}sica, Universidade Federal do Par\'{a}, Campus Universit\'{a}rio de Salin\'{o}polis,
 68721-000, Salin\'opolis, Par\'{a}, Brazil}

%=====================================================
%-----------------------------------------------------------------
\date{today; \LaTeX-ed \today}
%========================================================
\begin{abstract}
  We apply the Simpson-Visser phenomenological regularization method to a cylindrically   
  symmetric solution of the Einstein-Maxwell equations known as an inverted black hole. 
  In addition to analyzing some properties of thus regularized space-time, including the  
  Carter-Penrose diagrams, we show that this solution can be obtained from the 
  Einstein equations with a source combining a phantom scalar field with a nonzero
  self-interaction potential and a nonlinear magnetic field. A similar kind of source is 
  obtained for the cylindrical black bounce solution proposed by Lima \textit{et al.} as a regularized 
  version of Lemos's black string solution. Such sources are shown to be possible for a certain 
  class of cylindrically, planarly and toroidally symmetric metrics that includes the 
  regularized solutions under consideration. 
\end{abstract}
%========================================================
\maketitle
%========================================================

%========================================================
\section{Introduction}\label{S:intro}
%========================================================

  Space-time singularities are known to be an undesired though common feature of \GR\ (GR) and other
  classical theories of gravity. A natural hope that all singularities will be inevitably suppressed by 
  quantum gravity effects is made somewhat uncertain by the fact that different models and
  approaches of quantum gravity, being translated to the classical language, generally lead to quite  
  different results, see, e.g., 
  \cite{QG1,QG2,QG3,kelly20,achour20,alonso22,kunst20,ash20a,bambi13}, 
  see also a discussion in \cite{we20}. One can find among them \bh-white hole transitions  
  \cite{QG2, QG3, kelly20, achour20,alonso22}, scenarios where the spherical radius beyond a \bh\ 
  horizon tends to a constant positive value \cite{kunst20}, configurations having no horizons 
  at all \cite{bambi13}, etc.  Such a diversity evidently indicates a still unfinished situation
  in developing quantum gravity by now. 
  
  It is therefore natural that attempts to simulate the possible effects of quantum gravity in the
  framework of classical space-time, leaving aside the details of quantization, cause rather large 
  interest. The recent proposal of Simpson and Visser (SV) \cite{simp18} can be regarded as one
  of such attempts: they regularized the \Scw\ metric by replacing the spherical radius $r$ with 
  the nonzero expression $r(u) = \sqrt{u^2 + b^2}$ ($b > 0$ is a regularization parameter), 
  thus avoiding the \Scw\ singularity at $r=0$. With different values of $b$, the resulting geometry 
  can represent a \wh\ (if $b > 2m$, where $m$ is the \Scw\ mass), a \bh\ with two horizons 
  if $b < 2m$, and an intermediate case of an extremal \bh\ with a single horizon if $b =2m$.
  In the \bh\ case $b < 2m$, the sub-horizon region is a Kantowski-Sachs type cosmology in which
  $u$ is a temporal coordinate, and the hypersurface $u=0$, where the radius $r(u)$ 
  has a regular minimum, is actually a bounce of one of the two scale factors. This phenomenon 
  was termed a {\it black bounce} \cite{simp18}. One can also recall that such black bounces are 
  a common feature of one more class of space-time models, called {\it black universes}, 
  that is, \bhs\ in which the internal region is a Kantowski-Sachs cosmology with late-time 
  isotropization \cite{bk-uni05, bk-uni06, bk-uni12}. Such solutions have been obtained in GR
  with a phantom scalar field as well as in some scalar-tensor theories \cite{clem09, azreg11, trap18}.
  In the case $b=2m$, the hypersurface $u=0$ is null, being simultaneously a \bh\ horizon and 
  a throat, thus it may be called a {\it black throat} \cite{rahul22}. 
  
  Later on, a similar regularization was considered for the Reissner-Nordstr\"{o}m space-time 
  \cite{franzin21}. Lobo \textit{et al}. \cite{lobo20} constructed a large class of regular \bh\ and \wh\ 
  space-times. The diversity and richness of geometries obtained in this manner have attracted 
  much attention, and their rotating extensions and those with a NUT charge were also 
  obtained and studied \cite{mazza21, xu21, shaikh21, kam23, barr22}. 
  A further analysis concerned gravitational wave echoes at possible \bh/\wh\ transitions, 
  quasinormal modes and gravitational lensing in such space-times \cite{churilova19, yang21, 
  guerrero21, tsukamoto21, islam21, cheng21, kb20, tsukamoto20, lima21, nascimento20, franzin22}.
  
  A question of interest emerging in this connection is whether these new regular geometries can 
  be presented as solutions of GR with some reasonable sources, such as, for example, scalar 
  and/or \emag\ fields. As shown in \cite{rahul22} and \cite{canate22,rodrigues23}, a large class 
  of SV-like regular metrics, including the regularized \Scw\ and \RN\ metrics, can be viewed 
  as exact solutions to the Einstein equations with a sum of two stress-energy tensors (SETs): 
  that of a minimally coupled self-interacting phantom scalar field and that of a nonlinear
  \emag\ field in the framework of nonlinear electrodynamics (NED) with a Lagrangian as 
  a function of the invariant $F = F\mn F\MN$. A scalar field or NED taken separately cannot
  be such a source due to algebraic properties of their SETs. For regularized \Scw\ and 
  \RN\ solutions, the explicit forms of scalar and NED constructed were found as well as their 
  global structure diagrams for the cases of three and four horizons. Similar regularizations for 
  some cosmological models were considered in \cite{kam22}.
  
  In \cite{kb22s}, similar regularizations were obtained for two other families of singular \ssph\ 
  solutions of GR: Fisher's  solution with a massless scalar field \cite{fisher48} and a special subset 
  of dilatonic \bh\ solutions whose source consists of interacting scalar and \emag\ fields 
  \cite{dil1, dil2, dil3, dil4}. In both cases, the SV trick ($x \mapsto \sqrt{u^2 + b^2}$) is applied 
  in the simplest way to a factor $x$ producing a space-time singularity at its zero value. 
  Scalar-NED sources have also been found in \cite{kb22s} for regularized versions of these 
  space-times, and it turned out that the necessary scalar fields cannot be only canonical 
  (possessing positive kinetic energy) or only phantom (having negative kinetic energy), 
  but change their nature from one region to another in a regular manner. A situation where 
  a scalar is phantom in a strong-field region and canonical elsewhere has been termed a
  ``trapped ghost'' \cite{trap10}, and some globally regular \bh\ and \wh\ solutions with 
  such fields were obtained \cite{kroger04, trap10, don11}.  
  
  As a by-product, it has been shown in \cite{kb22s} that a sum of NED and minimally coupled scalar 
  SETs can provide a source for {\it any\/} \ssph\ metric, but in general the scalar can somewhere 
  change its nature from canonical to phantom. A way to describe an arbitrary \ssph\ metric using a 
  single field source was found in \cite{kb23} in the framework of the general 
  (Bergmann-Wagoner-Nordtvedt) scalar-tensor theory of gravity, but such a description turned out
  to be possible only piecewise. This was demonstrated using as examples the \RN\ metric and 
  the regularized \Scw\ metric according to \cite{simp18}.
  
  One more family of \bh\ space-times, those with cylindrical symmetry, called 
  black strings \cite{lemos95, cardoso01}, with the metric
\beq         \label{string}
		ds^2 = A(r) dt^2 - \frac {dr^2}{A(r)} - r^2 (dz^2 + d\varphi^2),\qq
		A(r) =  \alpha^2 r^2 - \frac b {\alpha r}, \qq \alpha, b = \const > 0,
\eeq   
  singular at $r=0$, was also regularized in the manner of \cite{simp18}, by replacing 
  $r \to \sqrt{u^2 + a^2}$, $a >0$, with, however, $dr \to du$ \cite{lima22}. 
  (Note that $r, u, a, \alpha^{-1}$ have here the dimension of length while $z$ and $b$ 
  are dimensionless.\footnote{We adopt the metric signature $(+,-,-,-)$ and use the geometric units where $8\pi G=c=1$.}) 	
  The resulting nonsingular metric has the form 
\beq         \label{string-reg}
		ds^2 = A(u) dt^2 - \frac {du^2} {A(u)} - (u^2+ a^2) (dz^2 + d\varphi^2),\qq
		A(u) =  \alpha^2 (u^2 + a^2) - \frac b {\alpha \sqrt{u^2 + a^2}}. 
\eeq     
  Both metrics \rf{string} and \rf{string-reg} are asymptotically AdS at large $r$ or $u$, with 
  the cosmological constant $\Lambda = - \alpha^2/3$.
  A further study of the metric \rf{string-reg} in \cite{lima22} concerned the energy conditions, 
  the thermodynamic properties of regular black strings, and possible stable and unstable 
  circular orbits of photons and massive particles.
  
  In the present paper, we study a similar regularization of another \scy\ metric of interest,
  called ``inverted \bhs'' and being a special solution to the Einstein-Maxwell equations
  \cite{kb79, Bronnikov:2019clf}. The metric has the form 
\beq       \label{ibh}
		ds^2 = A(x) dt^2 - \frac {dx^2}{A(x)} - \frac{q^2 x^2}{4k^2} (dz^2 + d\varphi^2),
		\qq       A(x) = \frac{16 k^4 (1-x)}{q^2 x^2},
\eeq  
  where $q$ (characterizing the electric or magnetic charge density) and $k$ are positive constants.
  The name ``inverted \bhs'' was proposed in \cite{kb79} because, contrary to ``normal'' \bhs, 
  a static region $x < 1$ in \rf{ibh} with a singularity at $x=0$ occurs at smaller values of the 
  circular radius $r (x) = qx/(2k)$, inside the horizon $x=1$, while the region $x>1$ 
  where $A < 0$ is nonstatic and represents a special kind of Bianchi-type I cosmology. 
  
  We regularize the metric \rf{ibh} as before, replacing $x \to \sqrt{u^2 + a^2}$, to obtain
\beq       \label{ibh-reg}
		ds^2 = A(u) dt^2 - \frac {du^2}{A(u)} 
					- \frac{q^2 (u^2+a^2)}{4k^2} (dz^2 + d\varphi^2),\qq  
			A(u) = \frac{16 k^4 (1-\sqrt{u^2+a^2})}{q^2 (u^2+a^2)},
\eeq  
  and study its properties in a manner similar to \cite{lima22}.  

  In addition to such regularizations, we construct the Carter-Penrose global structure diagrams 
  and determine possible field sources for both regularized metrics \rf{string-reg} and \rf{ibh-reg} 
  in the framework of GR similarly to \cite{rahul22} in terms of a combination of NED and 
  a self-interacting scalar field.
  
  It is necessary to mention that all the metrics \rf{string}--\rf{ibh-reg} can be interpreted 
  not only as \cy\ ones (such that $z \in \R$ and $\varphi\in [0, 2\pi)$) but also in terms 
  of two other kinds of symmetry: planar, if both $z \in \R$ and $\varphi\in \R$, and toroidal,  
  if both $z$ and $\varphi$ range on finite segments with identified ends. All local quantities
  discussed in this paper do not depend on these topological assumptions. We will adhere to
  interpretations in terms of cylindrical symmetry that seems to be of greater interest than the 
  other two.
    
  The paper is organized as follows. We begin the next Sec. II with giving some general  
  relations for \cy\ metrics and then consider some properties of the regularized metrics  
  \rf{string-reg} and \rf{ibh-reg}. Section III is devoted to finding and discussing field sources 
  for these metrics, and Sec. IV is a conclusion.
  
% =======================
\section{Regularized metrics} 
% =======================
\subsection{General relations}
% -------------------------------------------------

  In general, the line element that describes static space-times with cylindrical symmetry 
  is written as
\beq              \label{ds-cy}
    		ds^2=A(x)dt^2 - B(x)dx^2 - C(x)dz^2 - D(x)d\varphi^2.
\eeq
  However, in this paper we are dealing with space-times possessing locally flat orbits of the 
  spatial isometry group, such that $C(x) = D(x)$, and using the coordinate condition 
  $B(x) = 1/A(x)$, the metric can be written in the form\footnote
  		{In the whole paper, to avoid confusion, we denote by $r$ the quantity 
  		 $\sqrt{-g_{22}} = \sqrt{-g_{33}}$ 
  		 that  has a clear geometric meaning of a scale factor of 2-surfaces (cylinders) parametrized 
  		 by $x^2 = z$ and $x^3 = \varphi$ (similar to the spherical radius $r$ in the case of
  		 spherical symmetry). For the radial coordinate $x^1$ we use other letters, $x$ or $u$.} 
\beq               \label{ds}
		ds^2 = A(x) dt^2 - \frac {dx^2}{A(x)} - r^2(x) (dz^2 + d\varphi^2), 
\eeq  
  where we assume $z \in \R$ and $\varphi \in [0, 2\pi)$ according to cylindrical symmetry.
  The nonzero Riemann tensor components are
\bear               \label{Riem}
		K_1 = R^{01}{}_{01} = - \frac 12 A'', \qq\inch &&
		K_2 = R^{02}{}_{02} = R^{03}{}_{03} = - \frac{A'r'}{2 r},
\nn
		K_3 = R^{12}{}_{12} = R^{13}{}_{13} = - \frac{2Ar''+ A'r'}{2r}, \qq &&
		K_4 = R^{23}{}_{23} = - \frac{Ar'^2}{r^2},
\ear  
  where the coordinates are numbered as $(t,x,z,\varphi) = (0,1,2,3)$, 
  and primes denote $d/dx$.  Since the Kretschmann invariant 
  $\cK =R_{\mu\nu\rho\sigma}R^{\mu\nu\rho\sigma} = 4K_1^2+8K_2^2+8K_3^2+4K_4^2$
  is a sum of squares, it is clear that the finiteness of all $K_i$ from \rf{Riem} is a 
  {\it necessary and sufficient condition\/} for finiteness of all algebraic invariants of the 
  Riemann tensor. 
  
  Also, in full similarity with the more familiar case of spherical symmetry, regular 
  zeros of $A(x)$ (provided $r\ne 0$) correspond to Killing horizons at which the Killing vector 
  $\d_t$ that is timelike at $A > 0$, becomes null. At such horizons all $K_i$ are finite and 
  well-behaved, thus illustrating the well-known fact that Killing horizons are regular surfaces. 
  
  Let us also present the expressions for nonzero components of the Einstein tensor
  $G\mN = R\mN - \frac 12 \delta\mN R$, to be used in the Einstein equations 
  $G\mN = - T\mN$, where $T\mN$ is the stress-energy tensor of matter:
\bearr                 \label{Gmn}
		G^0_0 = \frac {1}{r^2} \Big[ A(2rr'' + r'^2) + A' rr' \Big],
\nnn 
		G^1_1 = \frac {1}{r^2} \Big[ A r'^2 + A' rr' \Big],				
\nnn
		G^2_2 = G^3_3 = \frac 1 r \Big[ Ar'' + A'r' + \frac 12 A'' r\Big].
\ear    
  
% -------------------------------------------------  
\subsection{The regularized black string}
% -------------------------------------------------

  Let us begin with the metrics \rf{string} and \rf{string-reg}, already considered in \cite{lima22}.
  In terms of \rf{ds}, in the metric \rf{string} we have 
\beq
		r(x) = x, \qq A(x) = \alpha^2 x^2 - b/(\alpha x). 
\eeq		
  Outside the horizon, at $x > x_h = b^{1/3}/\alpha$, there is a static (R-) region with an AdS-like 
  asymptotic behavior at large $x$. Inside the horizon, at $0 < x < x_h$, there is a dynamic (T-)
  region with a special Bianchi-type I geometry and a singularity at $x=0$. 
  
  The regularized metric \rf{string-reg}, with $u \in \R$, has the same AdS-like behavior at 
  large $r$ as the original one, but now this happens at both limits $u \to \pm \infty$ since
  $r = \sqrt{u^2+a^2}$.  
  
\begin{figure*}   
\centering   
\includegraphics[width=11cm]{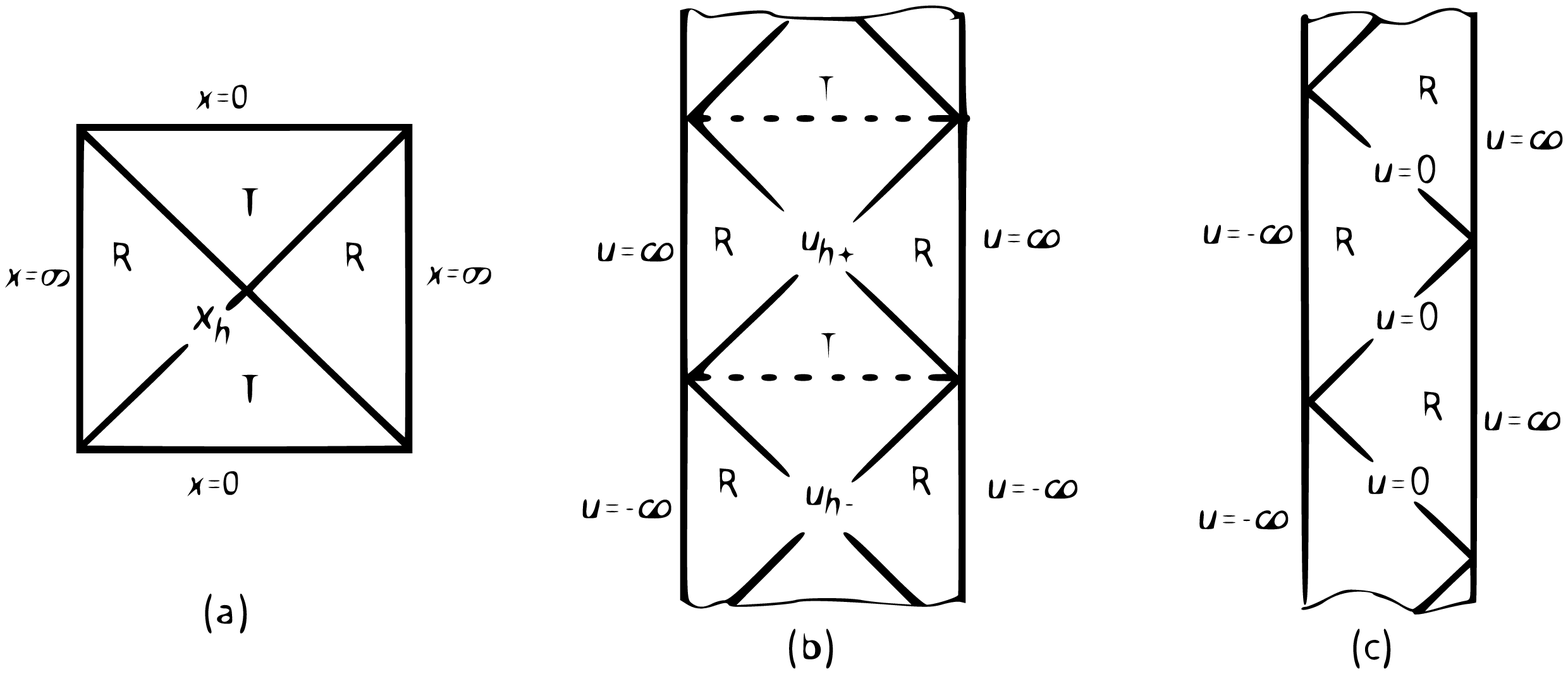}
\caption{\small
		Carter-Penrose diagrams for the black string metric \rf{string} [panel (a)] and its
		regularized version \rf{string-reg}: panel (b) for $a < x_h$, and panel (c) for $a =x_h$.
		All inner lines in the diagrams depict horizons, the letters R and T mark static (R-)
		and dynamic (T-) regions, respectively. The dashed line in panel (b) shows the 
		black bounce instant $u=0$. Diagrams (b) and (c) are infinitely continued 
		upward and downward.}   
  \label{Fig:Penr_String}
\end{figure*}

  The global properties of space-time \rf{string-reg} depend on the value of the 
  regularization parameter $a$:
\begin{itemize}  
\item
  		If $a < x_h$, it is a regular \bh, but now, unlike \rf{string} (see diagram (a) in 
  		Fig. \ref{Fig:Penr_String}), it has two horizons at $u = u_{h\pm} = \pm \sqrt{x_h^2-a^2}$
  		(diagram (b) in Fig. \ref{Fig:Penr_String}). At $u=0$ there is a minimum of $r(u)$ 
  		in a T-region,	in other words, a black bounce.
\item
		If $a = x_h$,  it is a regular \bh\ with a single extremal horizon at $u=0$, it 
		separates two R-regions (diagram (c) in Fig. \ref{Fig:Penr_String}).
\item
		If $a > x_h$, it is a \cy\ \wh\ with a throat at $u=0$.  
\end{itemize}
   
% -------------------------- fig 1
% --------------------------   
    
  The physical properties of space-time \rf{string-reg} are discussed in detail in \cite{lima22}.
   
% ------------------------------------------------------------
\subsection{The regularized inverted black hole}
% ------------------------------------------------------------

\begin{figure*}   
\centering   
\includegraphics[width=14cm]{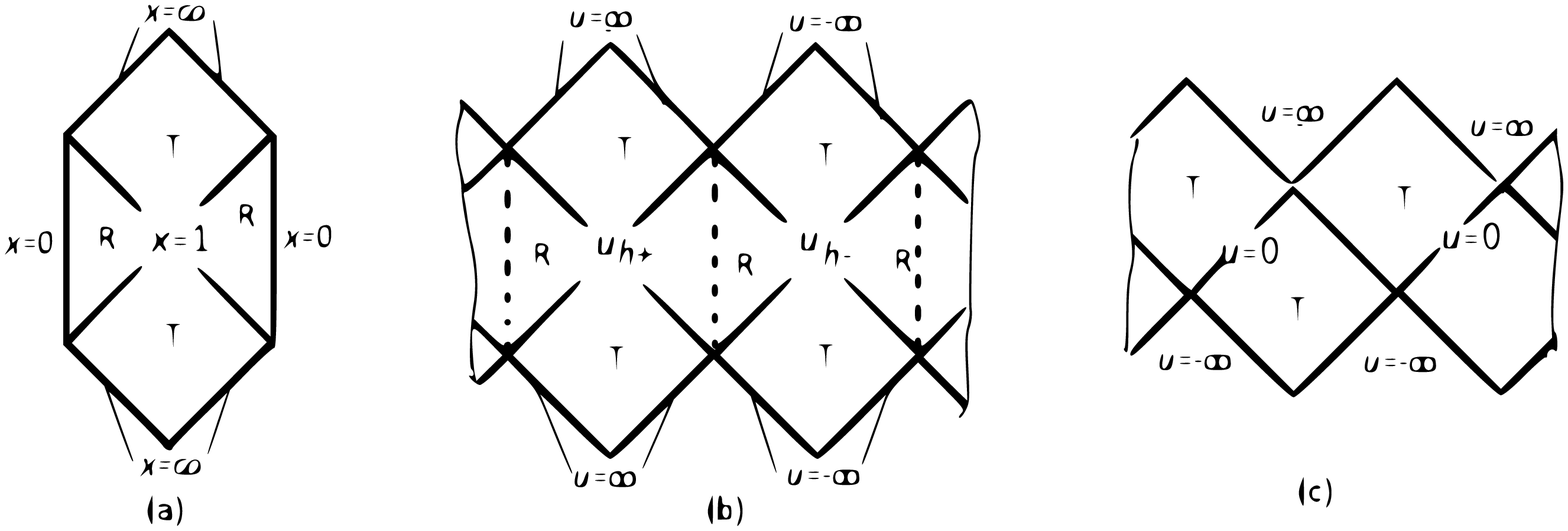}
\caption{\small
		Carter-Penrose diagrams for the inverted \bh\ metric \rf{ibh} [panel (a)] and its
		regularized version \rf{ibh-reg}: panel (b) for $a < 1$, and panel (c) for $a =1$.
		All inner solid lines in the diagrams depict horizons, the letters R and T mark static 
		(R-) and dynamic (T-) regions, respectively. The dashed line in panel (b) shows the 
		\wh\ throat $u=0$. Diagrams (b) and (c) are infinitely continued to the left and 
		to the right.} \label{Fig:Penr_Inv}
\end{figure*}

  The inverted black hole line element \rf{ibh} contains the metric functions in terms of \rf{ds} 
\beq         \label{Ar-ibh}
    		A (x) = \frac{16 (1-x)k^4}{q^2x^2}, \qq 
    		r^2(x) = \frac{q^2x^2}{4k^2}>0.
\eeq
  This space-time contains a horizon at $x=1$ and a singularity at $x=0$, where $r\to 0$
  (a singular axis in \cy\ space-time that may be interpreted as a charged thread repelling 
  neutral particles since $g_{00} = A(x) \to \infty$ there). 
  The horizon at $x=1$ has a common feature with a de Sitter horizon in that the metric 
  has a cosmological, time-dependent nature at larger values of $r$. 
  The ``far end'', $x\to \pm \infty$, corresponds to an infinitely remote past or 
  future in which the universe is highly anisotropic ($x$ is a time coordinate there, the 
  scale factor in the $t$ direction, now spatial, is $a_t \sim |x|^{-1/2}\to 0$, while the 
  other two scale factors are $a_z = a_\varphi \sim |x| \to \infty$).  
  The corresponding global causal structure diagram is presented in 
  Fig. \ref{Fig:Penr_Inv}, panel (a). 
  
  In the regularized metric \rf{ibh-reg}, we have, in terms of the new coordinate $u$, 
\beq         \label{Ar-ibh-reg}
		A (u) = \frac{16 k^4 \big(1-\sqrt{u^2 + a^2}\big)}{q^2(u^2 + a^2)}, \qq 
    		r^2(u) = \frac{q^2 (u^2 + a^2)}{4k^2}, \qq u \in \R.
\eeq  
  As was the case with the black string metric, the global properties of this space-time 
  depend on the value of the parameter $a$:
\begin{itemize}  
\item
  		If $a < 1$, we obtain a \cy\ \wh\ with a throat at $u=0$, surrounded by two 
  		cosmological-type horizons at $u = u_{h\pm} = \pm \sqrt{1 - a^2}$
  		(see diagram (b) in Fig.\,\ref{Fig:Penr_Inv}).
\item
		If $a = 1$,  it is a regular Bianchi-type I cosmological model with a single extremal 
		horizon at 	$u=0$ that separates two T-regions (diagram (c) in Fig.\,\ref{Fig:Penr_Inv}).
\item
		If $a > 1$, it is a regular Bianchi-type I cosmological model with a bounce 
		(minimum of $r$) at $u=0$.
\end{itemize}

% -------------------------- fig 2
% --------------------------   

  By substituting \rf{Ar-ibh-reg} to the expressions \rf{Riem} for $K_i$ it is straightforward 
  to verify that this metric is globally regular. Still we present, for completeness, the 
  corresponding expression for the Kretschmann scalar:
\beq
		\mathcal{K}=\frac{256 k^8 \left[-3 a^4 (u^2-4)-40 a^2 u^2+9 a^6+4 u^4 (3 u^2+14)\right]}
		{q^4 (a^2+u^2)^6}
		-\frac{1024 k^8 (-11 a^2 u^2+5 a^4+12 u^4)}{q^4 (a^2+u^2)^{11/2}},
\eeq
  where it is evident, in particular, that $\cK$ is finite at $u=0$ and decays as $u^{-6}$
  at large $|u|$.
  
  Substituting the quantities \rf{Ar-ibh-reg} to \rf{Gmn}, we find the components of the 
  Einstein tensor: 
\bear                 \label{G00-h}
		G^0_0 &=& - \frac{16 k^4 \big[2 a^2 (\sqrt{a^2 + u^2}-1) + u^2\big]}
							{q^2 (a^2 + u^2)^3},
\\               \label{G11-h}
		G^1_1 &=& - \frac{16 k^4 u^2}{q^2 (a^2 + u^2)^3},
\\                 \label{G22-h}
		G^2_2 &=& \frac{8 k^4 \big(2 u^2 - a^2 \sqrt{a^2 + u^2}\big)}{q^2 (a^2 + u^2)^3}.
\ear
  Using the Einstein equations, we can find the effective stress-energy tensor 
  that can be a material source of the metric, and analyze the fulfillment or violation of 
  various energy conditions by a source of metric under consideration. 
  
  It is, however, necessary to bear in mind that only in a static region ($A>0$)
  we deal with the usual relations
\beq
		\rho = - G^0_0, \qq  p_r = G^1_1, \qq p_\bot = G^2_2,
\eeq  
  where $\rho$ is the density, $p_r$ the radial pressure, and $p_\bot$ the tangential 
  pressure. In T-regions, where $A <0$, we must identify
\beq
		\rho = - G^1_1, \qq  p_r = G^0_0, \qq p_\bot = G^2_2.
\eeq   

\begin{figure}
\centering
\includegraphics[scale=0.7]{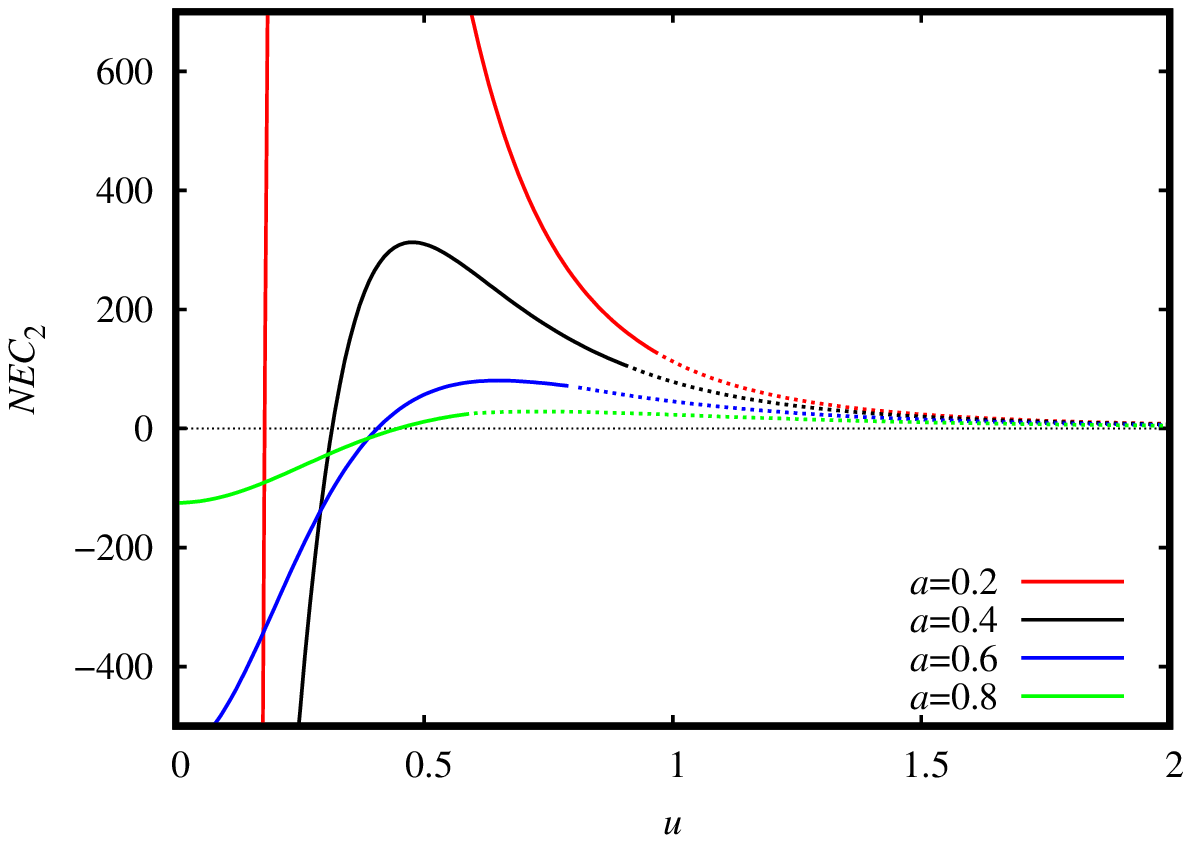}
\includegraphics[scale=0.7]{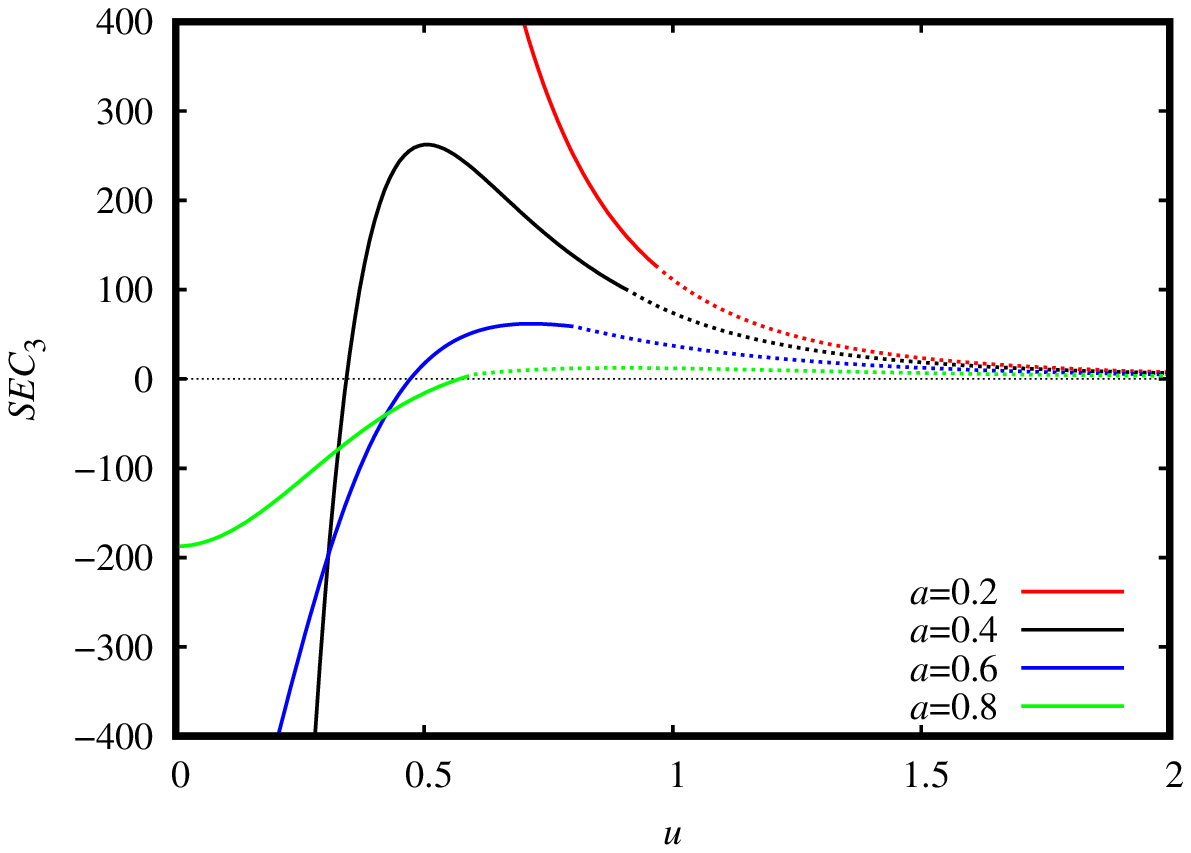}
\includegraphics[scale=0.7]{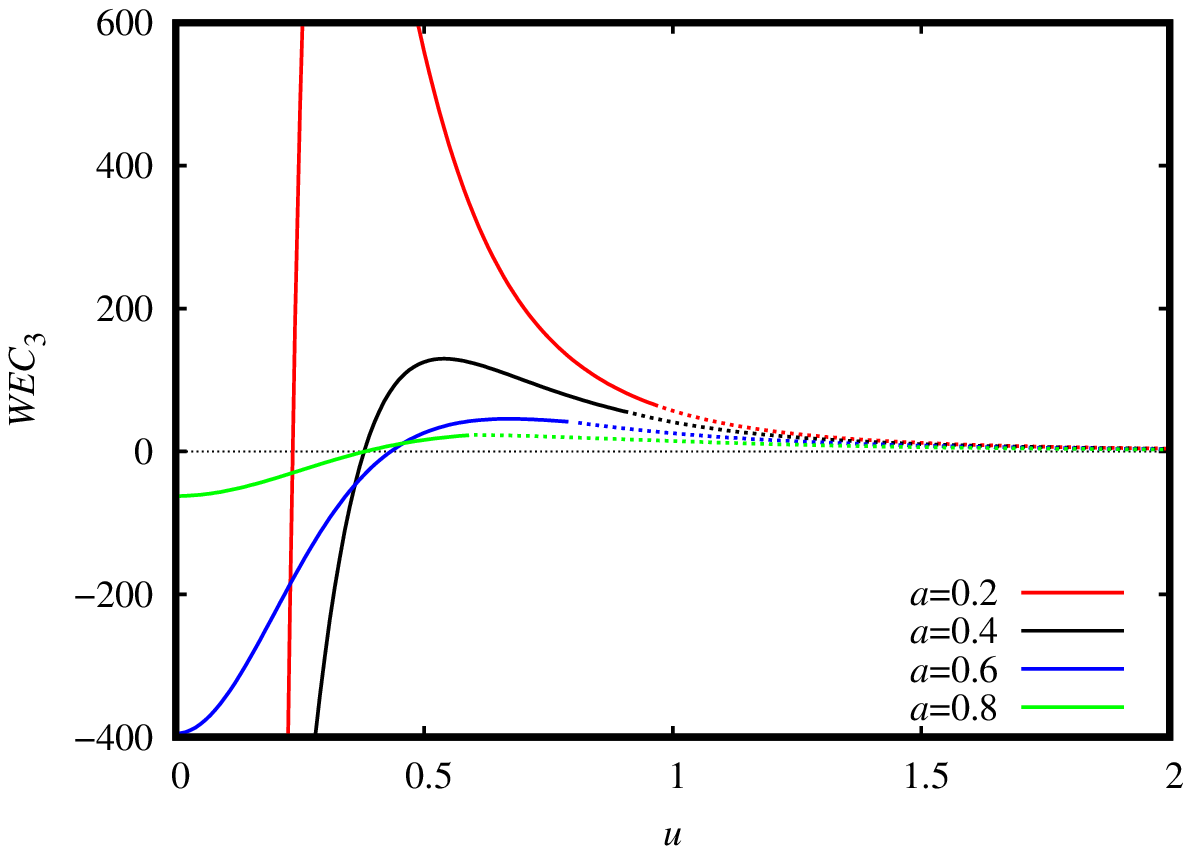}
\caption{Plots characterizing the fulfillment of energy conditions at $k=1$, $q=0.5$, and 
	    different values of $a$. Solid lines present the functions inside the horizon
	    (at $A > 0$), and dashed lines outside the horizon (at $A < 0$).}
    \label{fig:EC}
\end{figure}

\def\NEC{{\rm NEC}} 
\def\DEC{{\rm DEC}}
\def\SEC{{\rm SEC}}
\def\WEC{{\rm WEC}}
  As we are working with solutions containing throats or bounces that require exotic sources, 
  it is of interest to analyze the fulfillment of the 
  null (NEC), weak (WEC), strong (SEC) and dominant (DEC) energy conditions:
  They can be presented as follows:
\bear
	&&  \NEC_{1,2}= \WEC_{1,2}= \SEC_{1,2}
		\Longleftrightarrow \rho+p_{r,\bot} \geq 0, 					\label{Econd1} 
\\
	&&  \SEC_3 \Longleftrightarrow  \rho+p_r+2p_\bot\geq  0,		\label{Econd2}
\\
	&&  \DEC_{1,2} \Longrightarrow \rho - p_{r,\bot}\geq 0 
	\quad \mbox{and} \quad \rho+p_{r,\bot}\geq 0,					\label{Econd3}
\\
	&&  \DEC_3 = \WEC_3 \Longleftrightarrow  \rho\geq 0.			\label{Econd4}
\ear
  In essence, all energy conditions are violated once the NEC is violated, and in our case
  the condition $\NEC_1$ is everywhere violated once $r''/r>0$. The condition 
  $\DEC_1$ is also violated since it includes $\NEC_1$ as its component.
  In Fig.\,\ref{fig:EC} we illustrate the behavior of the other energy conditions. 
  At small values of $a$, these conditions are satisfied outside the horizon 
  (i.e., where $A < 0$).

  In these and all subsequent figures we restrict ourselves to positive values of $u$;
  for $u < 0$ the plots are unnecessary due to the symmetry $u \longleftrightarrow -u$.
% ------------------------------------ fig 3
% -----------------------------------------
\bigskip

% ====================================
\section{Field sources for the regularized metrics}
% ====================================
\subsection{General consideration}
% ------------------------------------------

  By analogy with \cite{rahul22, canate22}, let us find field sources for the \cy\ metrics 
  \rf{string-reg} and \rf{ibh-reg} in GR as NED plus 
  a self-interacting scalar field, with the total action
\beq
    S=\int \sqrt{-g}d^4x\left[R + 2h(\phi) g^{\mu\nu}\d_\mu \phi\d_\nu \phi
    		-2V(\phi) - L(F)\right],    			\label{Action}
\eeq
  where $g$ is the determinant of the metric $g\mn$, $\phi$ is the scalar field, $V(\phi)$ is its
  potential, $L(F)$ is the NED Lagrangian, 
  $F=F^{\mu\nu}F_{\mu\nu}$, and $F\mn = \d_\mu A_\nu - \d_\nu A_\mu$ is the 
  electromagnetic field tensor. The function $h(\phi)$ is included here not only for generality
  (expressing the freedom of scalar field parametrization), but also to include possible cases
  where $\phi$ changes its nature from canonical ($h>0$) to phantom ($h < 0$), exhibiting 
  a ``trapped ghost'' behavior \cite{trap10, kb22s}.
  
  Varying the action \eqref{Action} with respect to $\phi$, $A_\mu$, and $g^{\mu\nu}$, 
  we obtain the field equations
\bearr
		\nabla_\mu \left[L_F F^{\mu\nu}\right]
			=\frac{1}{\sqrt{-g}}\d_\mu \left[\sqrt{-g}L_F F^{\mu\nu}\right]=0,
														\label{eq-F}
\\  \lal
      2h(\phi) \nabla_\mu \nabla^\mu\phi + \frac{dh}{d\phi}\d^\mu\phi \d_\mu\phi
		+  \frac{dV(\phi)}{d\phi} =0,                      \label{eq-phi}
\\  \lal
      G\mN = -T\mN [\phi] - T\mN [F],				\label{eq-Ein}
\ear
  where $L_F=d L/d F$, $T\mN[\phi]$ and $T\mN[F]$ are the stress-energy tensors of the 
  scalar and electromagnetic fields, respectively:
\bearr                   \label{SET-F}
    T\mN [F] = \frac 12 \delta^{\mu}_\nu L(F ) -2 L_F {F}^{\nu\alpha} F_{\mu \alpha},
\yyy                     \label{SET-phi}
    T\mN[\phi] =  2 h(\phi) \d^\nu\phi\d_\mu\phi 
				    - \delta\mN \big (h(\phi) \d^\alpha \phi \d_\alpha \phi - V(\phi)\big).
\ear

  Let us show that any \scy\ metric of the form \rf{ds} can be presented as a solution 
  to \eqs \rf{eq-F}--\rf{eq-Ein} with a proper choice of $\phi$ and $F\mn$, quite similarly to
  the \ssph\ case studied in \cite{kb22s}. 
  The symmetry of the metric \rf{ds} makes us assume $\phi = \phi(x)$ and 
  single out among $F\mn$ either a ``radial'' electric field with the only nonzero component
  $F_{01}= -F_{10}$ or a ``radial'' magnetic field with $F_{23} =-F_{32}\ne 0$. 
  For certainty, let us consider a magnetic field, so that 
  $F_{\theta\varphi} = -F_{\varphi\theta}= Q= \const$, and $Q$ may
  be interpreted as a magnetic charge, while \eq \rf{eq-F} is trivially satisfied. Then the 
  \emag\ invariant $F$ is given by $F = 2Q^2/r^4$, and the stress-energy tensors
  \rf{SET-phi} and \rf{SET-F} take the form
\bearr         \label{T-phi}
		T\mN[\phi] = h(\phi) A(x) \phi'{}^2 \diag (1, -1, 1, 1) + \delta \mN V(\phi), 
\yyy           \label{T-F}		
		T\mN[F] = \frac 12 \diag\Big(L,\ L,\ L - \frac{4Q^2}{r^4} L_F,\
						L - \frac{4Q^2}{r^4} L_F\Big). 
\ear  

  For an arbitrary metric \rf{ds}, all three components \rf{Gmn} of the Einstein tensor are 
  different; therefore, taken separately, a scalar or \emag\ field cannot solve the problem
  because of the equalities $T^0_0 [F] = T^1_1 [F]$ for the \emag\ field and 
  $T^0_0 [\phi] = T^2_2 [\phi]$ for the scalar field. So we will consider them together.
  
  Assuming that $A(x)$ and $r(x)$ are known and substituting them to \eqs \rf{eq-Ein}, 
  we find
\beq              \label{T02}
  		G^2_2 - G^0_0 = T^0_0 - T^2_2 = \frac{2Q^2}{r^4} L_F ,
\eeq
  (the scalar field does not contribute to this combination of \eqs\rf{eq-Ein}). 
  Hence we know $L_F$ as a function of $x$, and recalling that $F(x) = 2Q^2/r^4(x)$, 
  is also known, we can write from \rf{T02}
\beq
			L' = \frac {F'}{F} (G^2_2 - G^0_0) = - \frac {4r'}{r} (G^2_2 - G^0_0), 
\eeq   
  and integrate. We thus know $L(x)$, $F(x)$ and finally $L(F)$, at least in 
  a range where $r(x)$ is monotonic.
  
  Furthermore, from \rf{eq-Ein} we have
\beq              \label{T01}
		T^0_0 - T^1_1 = 2 h(\phi) A(x) \phi'{}^2 = G^1_1 - G^0_0 = -2A(x) \frac{r''}{r},
\eeq   
  which gives us $h(\phi)\phi'{}^2$ as a function of $x$. 
  It is easy to notice that we have $h(\phi)\geq 0$ (the scalar field is canonical) as long
  as $r'' \leq 0$, and $h(\phi)\leq 0$ (the scalar field is phantom) if $r'' \geq 0$. 
  In the general case, $r''(x)$ can change its sign, and then the sign of $h(\phi)$ also varies. 
  In all cases we can freely choose the scalar field parametrization and conclude that we 
  know the functions $\phi(x)$ and $h(x)$. The only still unknown quantity, the potential 
  $V(\phi)$, can now be found from any components of \eqs \rf{eq-Ein}, for example, 
  $G^t_t = - T^t_t$. The scalar field equation \rf{eq-phi} is known to follow from the 
  Einstein equations \rf{eq-Ein} (actually, as long as $\phi \ne \const$, \eq \rf{eq-phi}
  directly follows from the ``conservation'' law $\nabla_\nu T^\nu_\mu$, which is in turn a
  consequence of the Einstein equations). So we can assert that the whole set of 
  equations is fulfilled. 
  
  Another algorithm consists in using again \rf{T01} to determine $\phi$ and $h(\phi)$
  and then finding $V(\phi(x))$ from the scalar field equation \rf{eq-phi}.
  With known $V(x)$, the function $L(F(x))$ is found from one more combination of 
  the Einstein equations,
\beq                 \label{T01+}            
		L(x) = -2V(x) - G^0_0 - G^1_1,
\eeq  
  and with known $F(x)$ it is then easy to find $L(F)$. Needless to say that both 
  algorithms must lead to the same result, up to the choice of integration constants when
  finding $L(x)$ in the first algorithm and $V(x)$ in the second one.  

  For our particular regular metrics \rf{string-reg} and \rf{ibh-reg}, $r'' > 0$, hence we will 
  inevitably deal with a phantom scalar field and can safely put $h(\phi) =-1$.
  
  We can note that a representation of any metric \rf{ds} using an electric instead of 
  magnetic field is also possible but is slightly more complicated.

% ----------------------------------------------------------------------------
\subsection{A field source for a regularized black string} 
% ----------------------------------------------------------------------------

  According to the above, since $r'' > 0$ in the metric \rf{string-reg}, we put 
  $h(\phi) = -1$. Let us try to model the solution with a magnetically charged source 
  and a phantom scalar field. The magnetic field and the invariant $F$ are given by
\beq   \label{F23bounce}
    			F_{23} = Q,\qq    F(u)=\frac{2Q^2}{(u^2+a^2)^2}.
\eeq

\begin{figure}
\centering
    \subfigure[]{\includegraphics[scale=0.7]{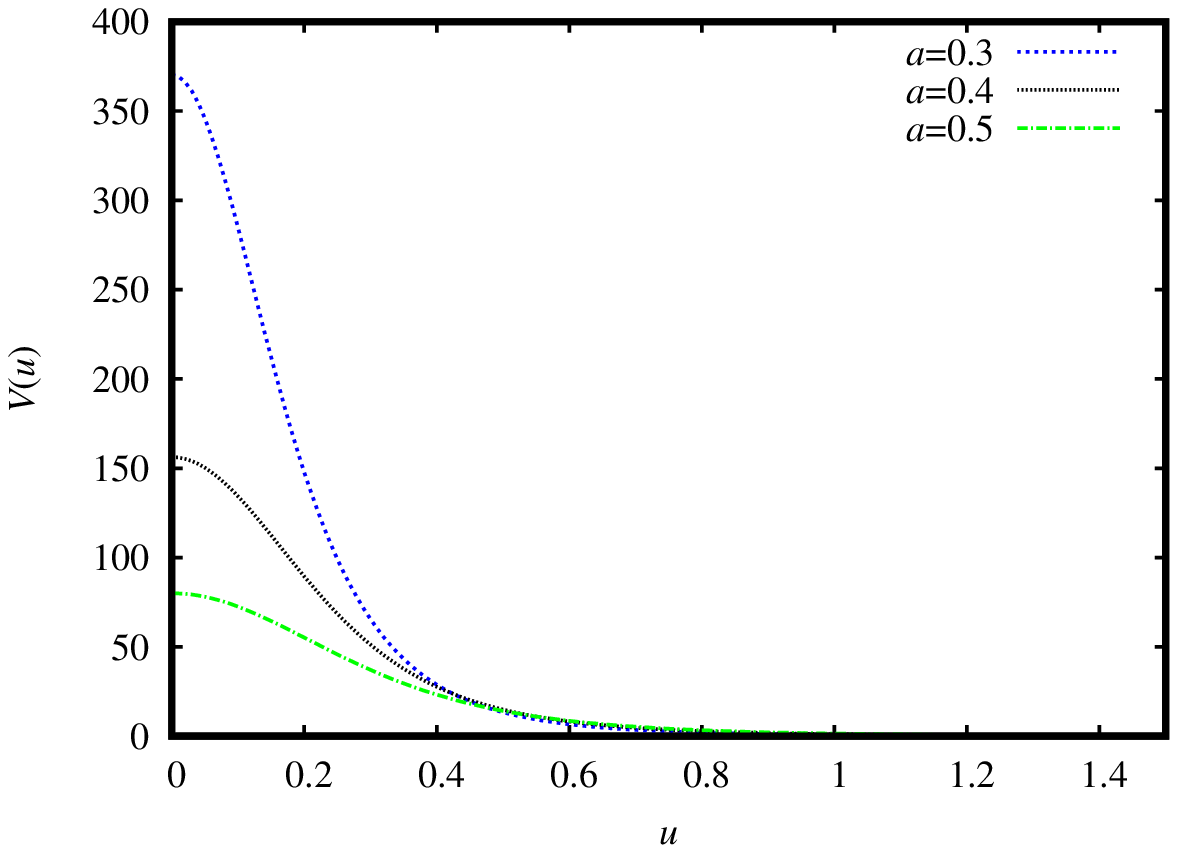}}
    \subfigure[]{\includegraphics[scale=0.7]{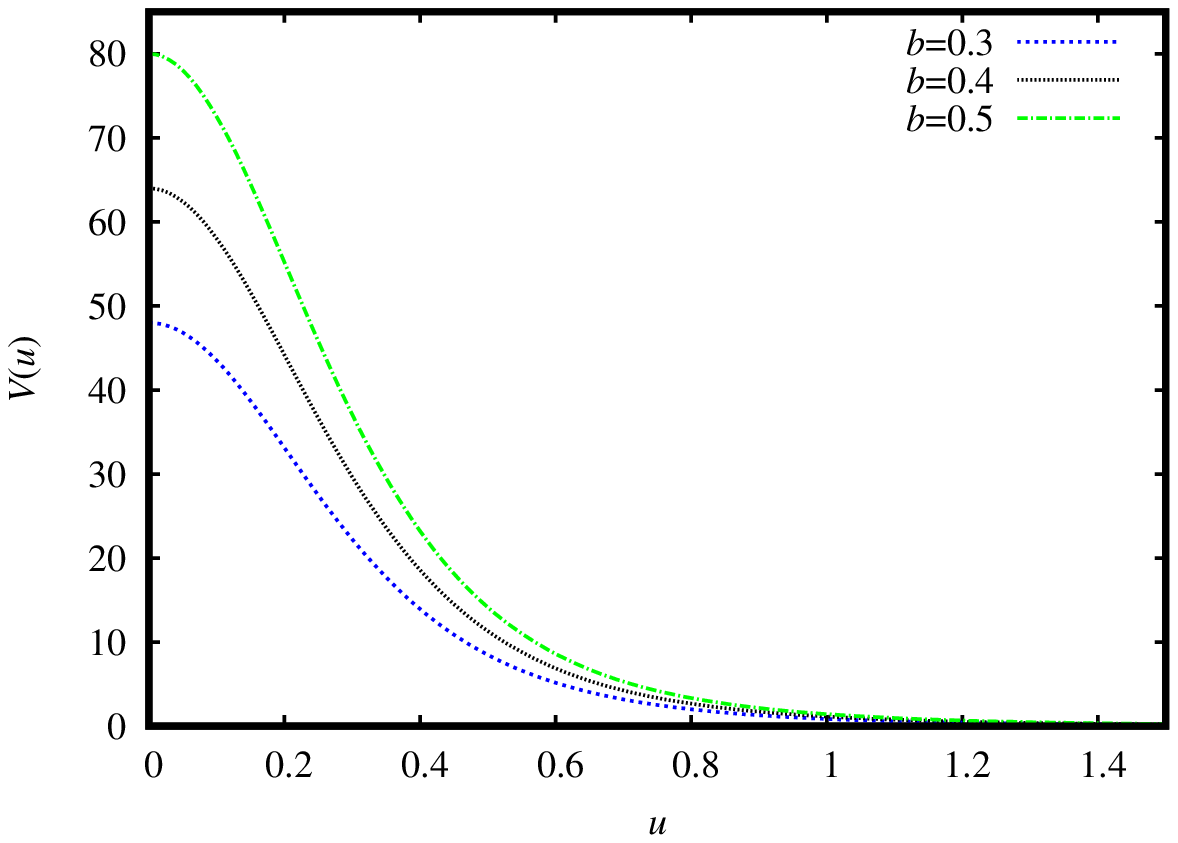}}
    \subfigure[]{\includegraphics[scale=0.7]{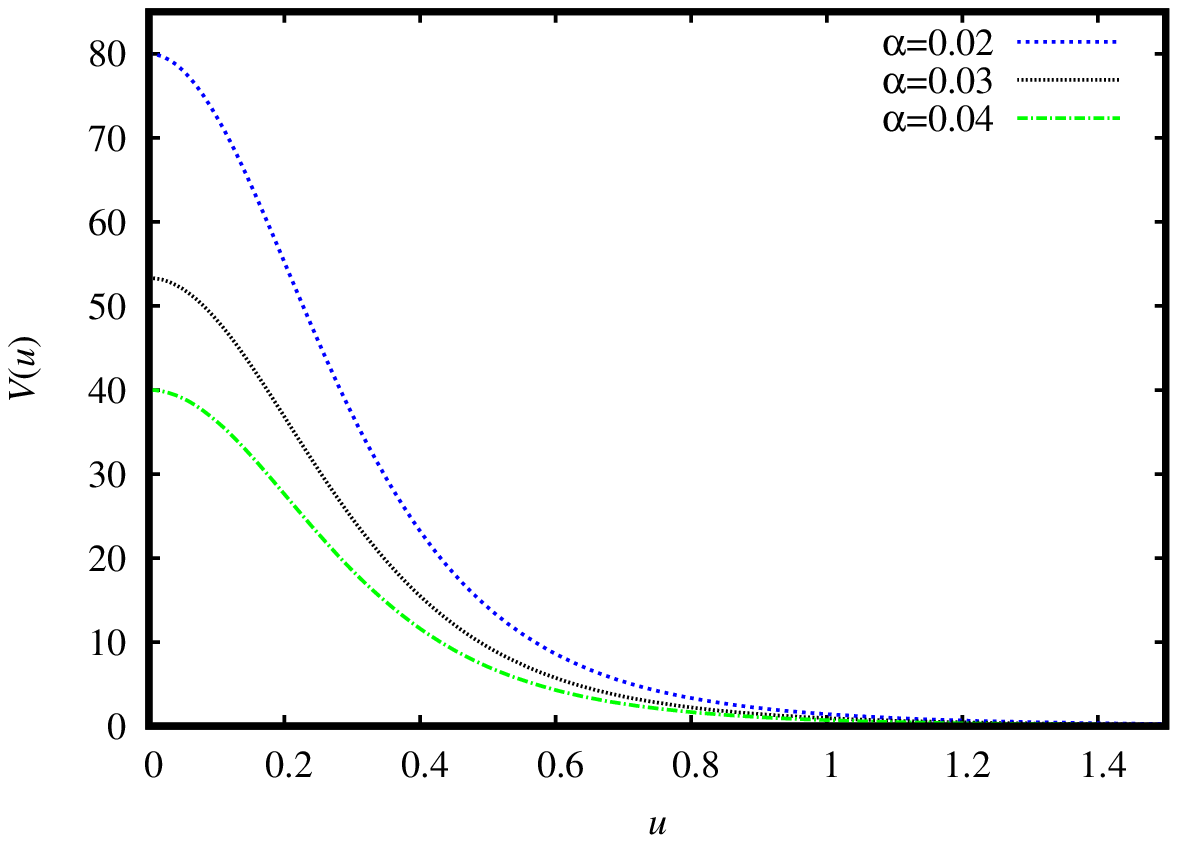}}
\caption{The potential $V(\phi(u))$, Eq. \rf{VuString}, as a function of the coordinate $u$.
    In (a) we fixed $b=0.5$ and $\alpha=0.02$. In (b) we fixed $a=0.5$ and $\alpha=0.02$. 
    In (c) we fixed $a=0.5$ and $b=0.5$.}
    \label{fig:VuString}
\end{figure}

\begin{figure}
\centering
    \subfigure[]{\includegraphics[scale=0.7]{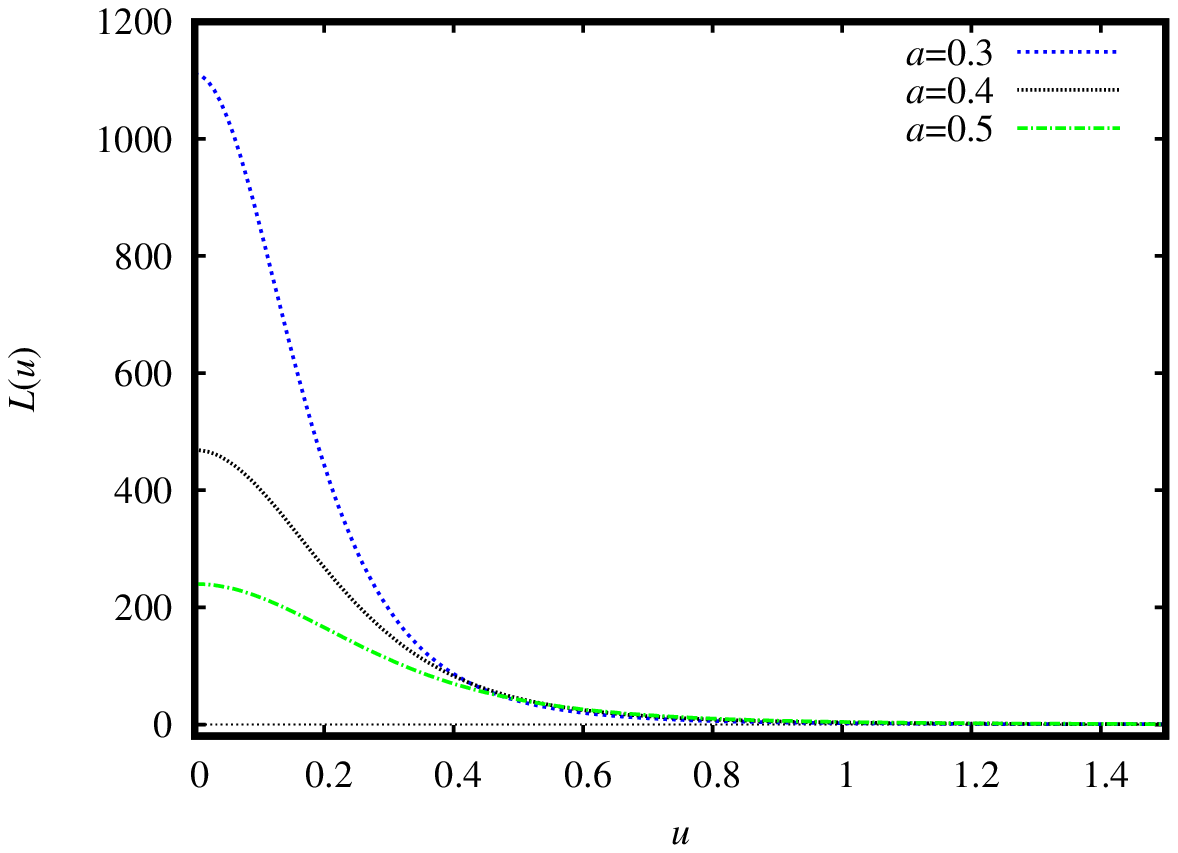}}
    \subfigure[]{\includegraphics[scale=0.7]{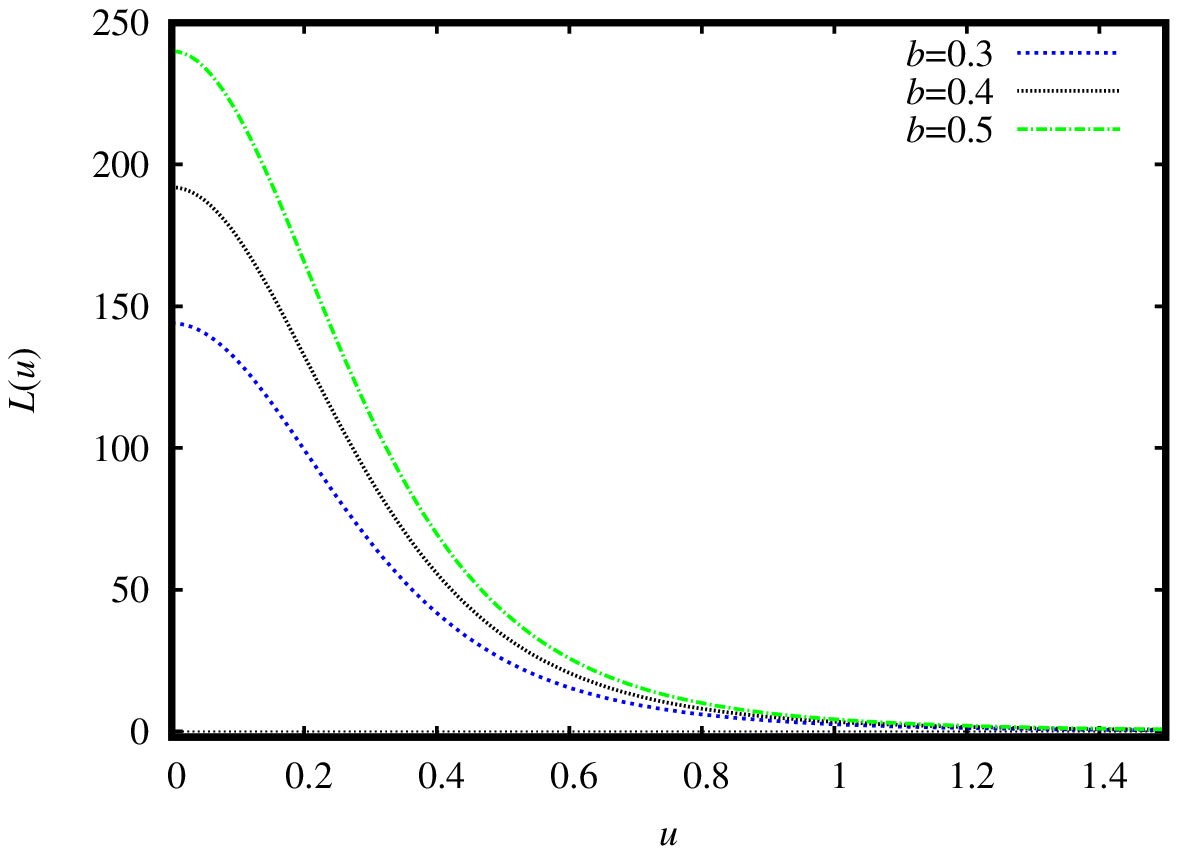}}
    \subfigure[]{\includegraphics[scale=0.7]{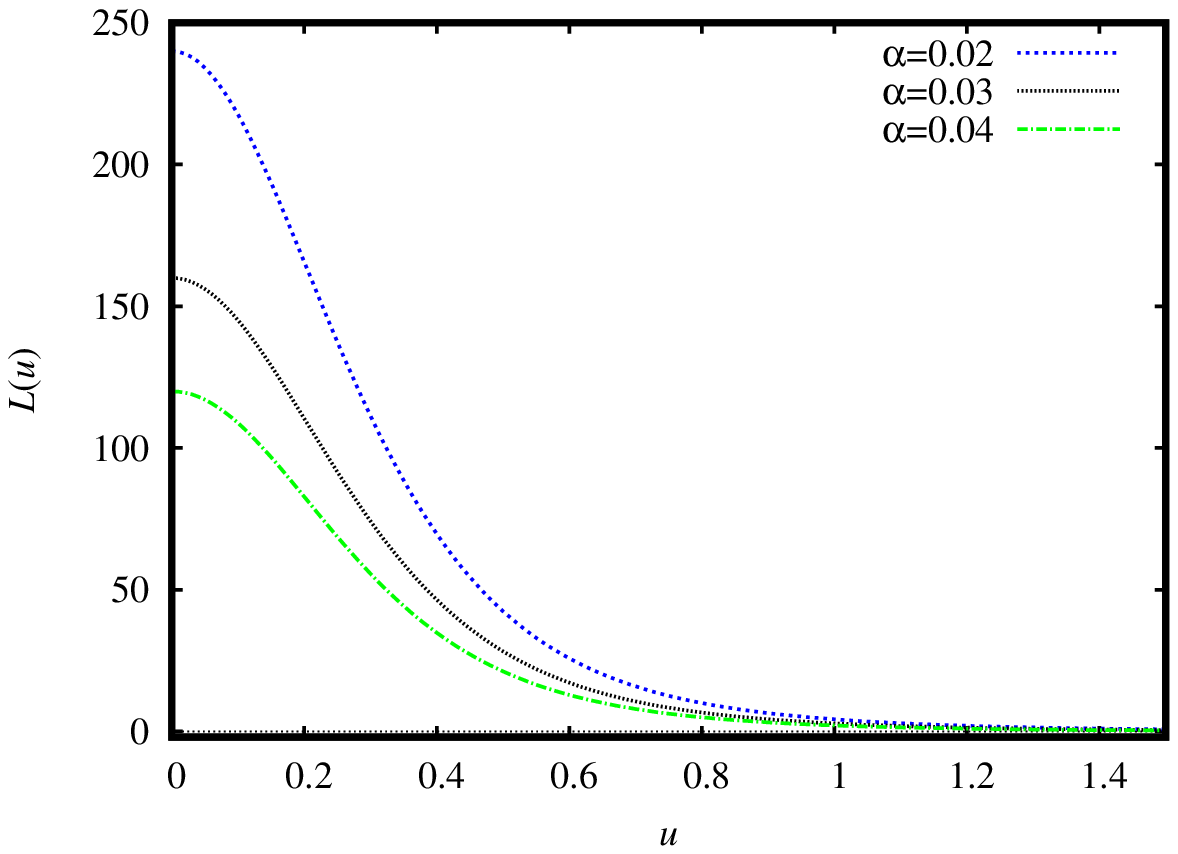}}
\caption{The function $L(F(u))$, Eq. \rf{LuString}, as a function of the coordinate $u$.
	In (a) we fixed $b=0.5$ and $\alpha=0.02$. In (b) we fixed $a=0.5$ and $\alpha=0.02$. 
	In (c) we fixed $a=0.5$ and $b=0.5$. At large $|u|$, the function $L(F(u))$ tends 
	to $- 6\alpha^2$ instead of zero, according to \eqn{LuString}, but this contribution
	is too small to 	be visible in the figure.	}
    \label{fig:LuString}
\end{figure}

  Let us use the second algorithm described above. From \rf{T01} we find
\beq
               \phi(u) = \arctan (u/a),
\eeq  
  and from the scalar equation \eqref{eq-phi} we determine,
  choosing the integration constant so that $V$ vanishes at large $|u|$,
\beq
   	 V(u) = 2a^2 \left(\frac{b}{5 \alpha  (a^2+u^2)^{5/2}}
    						+\frac{\alpha ^2}{a^2+u^2}\right).               \label{VuString}
\eeq
   Next, \eq \rf{T01+} yields
\beq
    L(u) = \frac{6 a^2 b}{5 \alpha  (u^2+a^2)^{5/2}}- 6 \alpha^2,  \label{LuString}
\eeq
   while the combination \rf{T02} leads to
\beq   
    		L_F(u) = \frac{3 a^2 b}{4 \alpha Q^2 \sqrt{u^2+a^2}},
\eeq
    These quantities satisfy the condition 
\beq           \label{corr}
   		 L_F - \frac{dL}{du}\left(\frac{dF}{du}\right)^{-1}=0,    
\eeq
  confirming the correctness of our calculations. We can now write $L(F)$ and $V(\phi)$ as
\bear             \label{LF-string}
    L(F) &= & \frac{6 a^2 b F}{5 \alpha} \bigg(\frac F{2Q^2}\bigg)^{5/4}  -6 \alpha ^2,
\\
    V(\phi)  & = &   2 \alpha^2 \cos^2\phi + \frac{2b}{5\alpha a^3}\cos^5 \phi.
\ear
 Their behavior as functions of $u$ is illustrated in Figs.\,\ref{fig:VuString} and \ref{fig:LuString}. 
 The magnitude of $V(u)$ and $L(u)$ decreases as $a$ and $\alpha$ increase, and 
 increases as $b$ increases. 
% ----------------------------------------- fig 4
% ------------------------------------------ fig 5

% ------------------------------------------------------------------------------
\subsection{A field source for a regularized inverted black hole} 
% ------------------------------------------------------------------------------

  Again, for a magnetically charged source, the magnetic field is given by
\beq        \label{F23ibh}
    			F_{23} = Q,\qq    F(u) = \frac {2Q^2}{r^4} 
    					 = \frac{32 Q^2 k^4}{q^4 (u^2+a^2)^2}.
\eeq
  Let us note that the magnetic charge $Q$ of our anticipated source 
  has nothing to do with the initial charge $q$ belonging to the ``seed'' solution 
  with the metric \rf{ibh} and involved in the regularized metric \rf{ibh-reg}.
  However, as a special case, they can coincide.
  
  As before, with $h(\phi) = -1$, for the scalar field we obtain from \rf{T01}
\beq
               \phi(u) = \arctan (u/a).
\eeq  
  Then the scalar field equation \rf{eq-phi} yields
\beq \label{VuInv}
		V(u) = \frac{32 a^2 k^4}{15 q^2 (u^2+a^2)^3} 
					\left(-5 + 3 \sqrt{u^2+a^2}\right).
\eeq    
  From the Einstein equations we then obtain 
\bear
      L(u)  &=& \frac{32 k^4}{15 q^2 (u^2+a^2)^3}
      			 \left[15 (u^2+a^2) + 9a^2 \sqrt{u^2+a^2} - 20 a^2\right],\label{LuInv}
\\      			 
   	 L_F(u) &=& \frac{q^2}{Q^2} \bigg(\frac{3a^2}{4 \sqrt{a^2+u^2}}
   	 				+\frac{u^2-a^2}{u^2+a^2}\bigg).
\ear   	 				
  As in the previous case, the correctness of calculations is verified by \eq \rf{corr}.
  With our magnetic source, the knowledge of $u(F)$ and $u(\phi)$ enables us to write 
  the Lagrangian $L(F)$ and the potential $V(\phi)$ as follows:
\bear
    L(F) &=& \frac{q^2 F}{Q^2} \bigg(1 + \frac{3 a^2}{5 x} - \frac{4a^2}{3 x^2}\bigg), 
    \qq     x = \frac{2k}{q}\bigg(\frac{2Q^2}{F}\bigg)^{1/4},        \label{LFInv}
\\
    V(\phi) &=& \frac{32 k^4 }{15 q^2 a^4} \cos^5\phi (3 a - 5\cos\phi).      \label{VphiInv}
\ear    
  There is good reason to suppose the equality $Q=q$ since it provides $L\approx F$
  as $F\to 0$, the asymptotically Maxwell behavior of $L(F)$ at small magnetic fields.
  So in what follows we put $Q=q$.

\begin{figure}
\centering
    \subfigure[]{\includegraphics[scale=0.7]{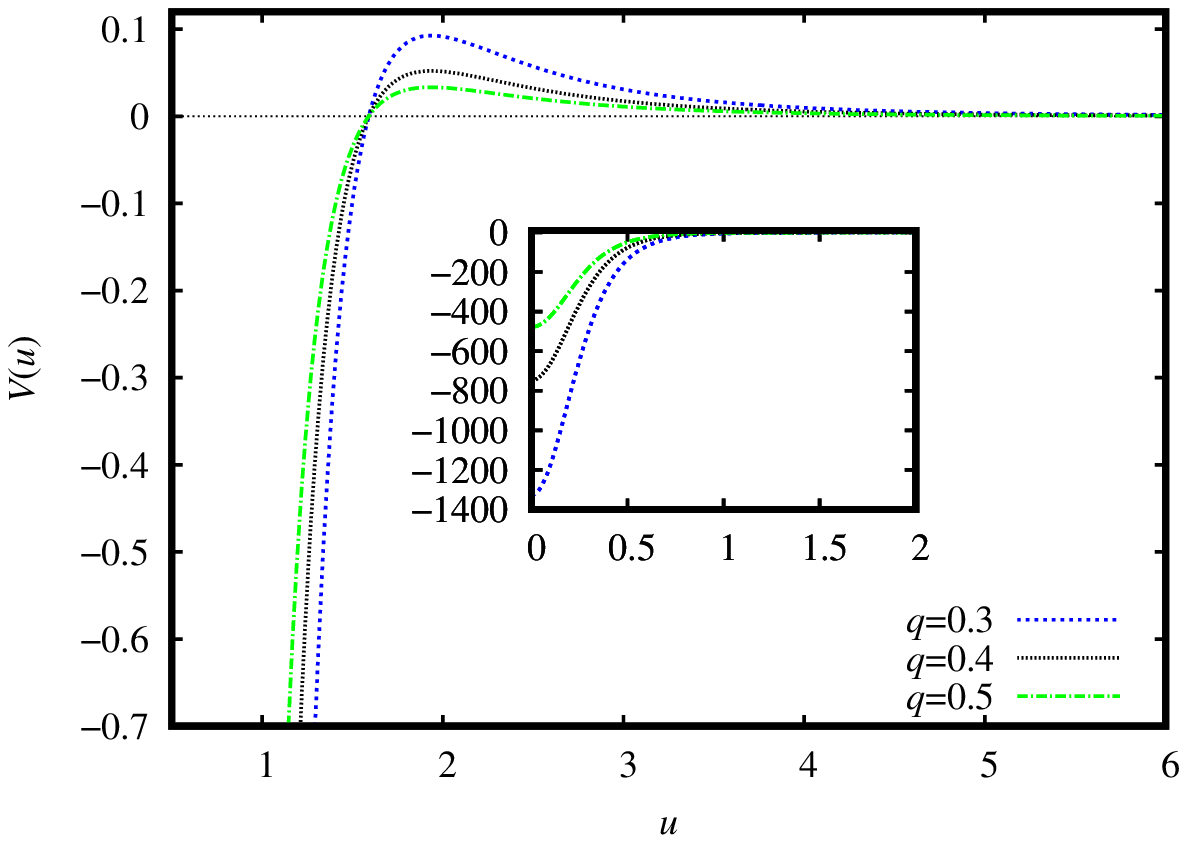}}
    \subfigure[]{\includegraphics[scale=0.7]{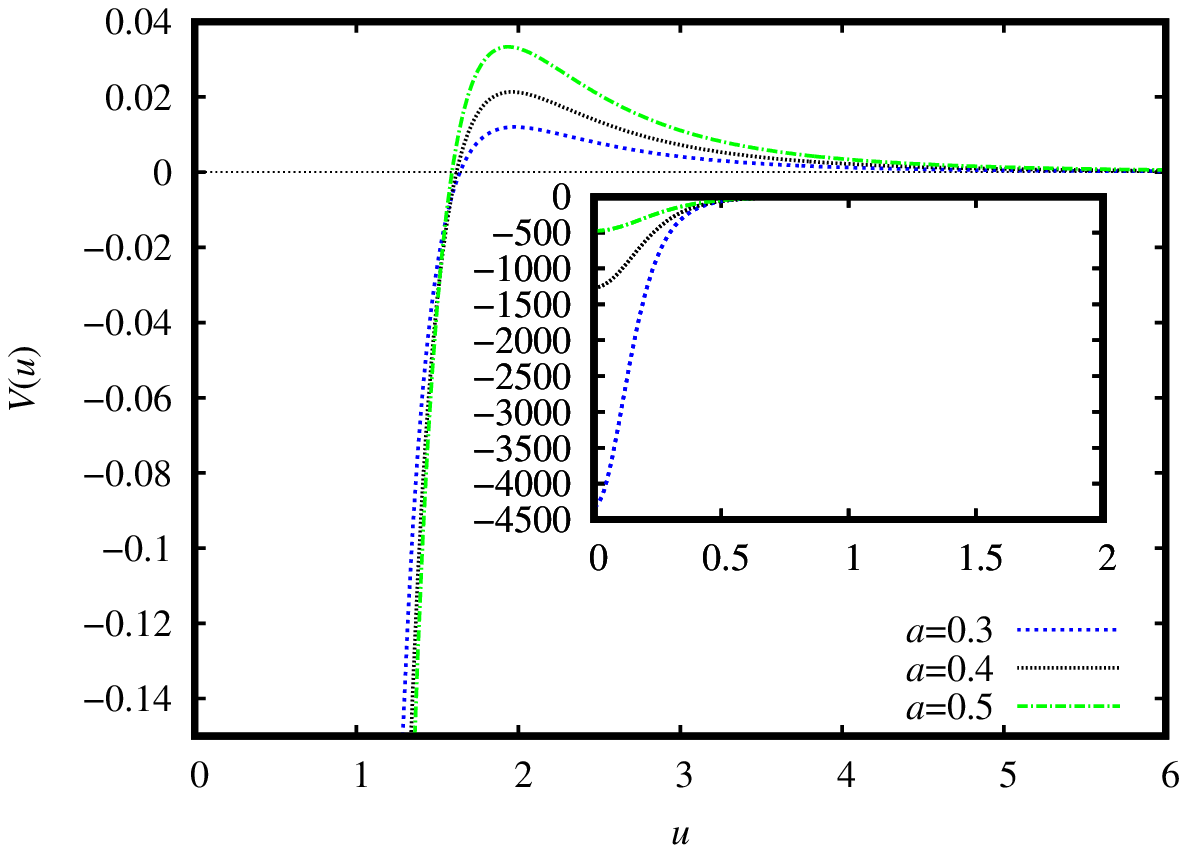}}
\caption{The potential $V(\phi(u))$, Eq. \rf{VuInv}, with $k=1$, as a function of the
    coordinate $u$. In (a) we fixed $a=0.5$ and in (b) we fixed $q=0.5$.}
    \label{fig:VuInverse}
\end{figure}

  The behavior of $V(u)$ is shown in Fig.\,\ref{fig:VuInverse}. With other values of 
  the parameters, the potential behaves qualitatively in the same way. As the 
  charge $q$ increases, the peaks of the potential become lower, while they grow at increasing 
  values of the parameter $a$.
% ------------------------------------------------ fig 6
% -------------------------------------------- fig 7
\begin{figure}
    \centering
    \subfigure[]{\includegraphics[scale=0.7]{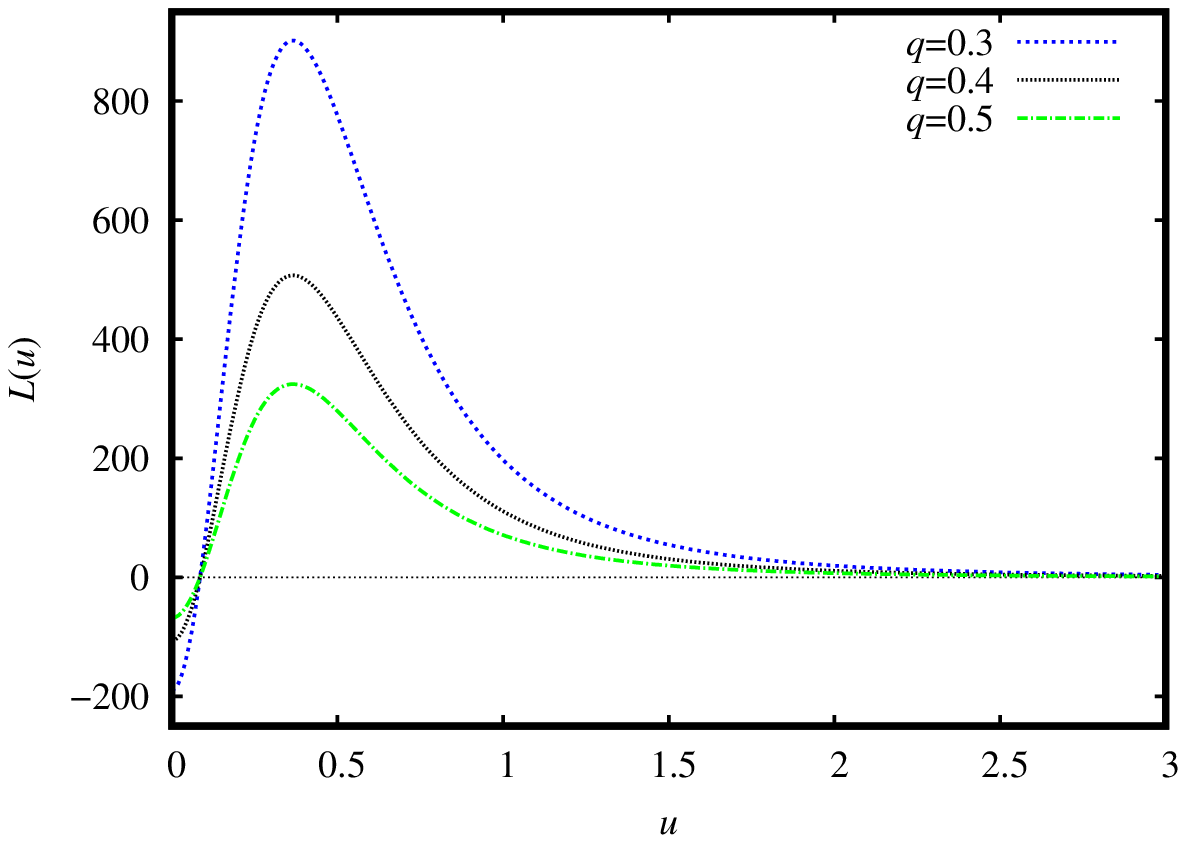}}
    \subfigure[]{\includegraphics[scale=0.7]{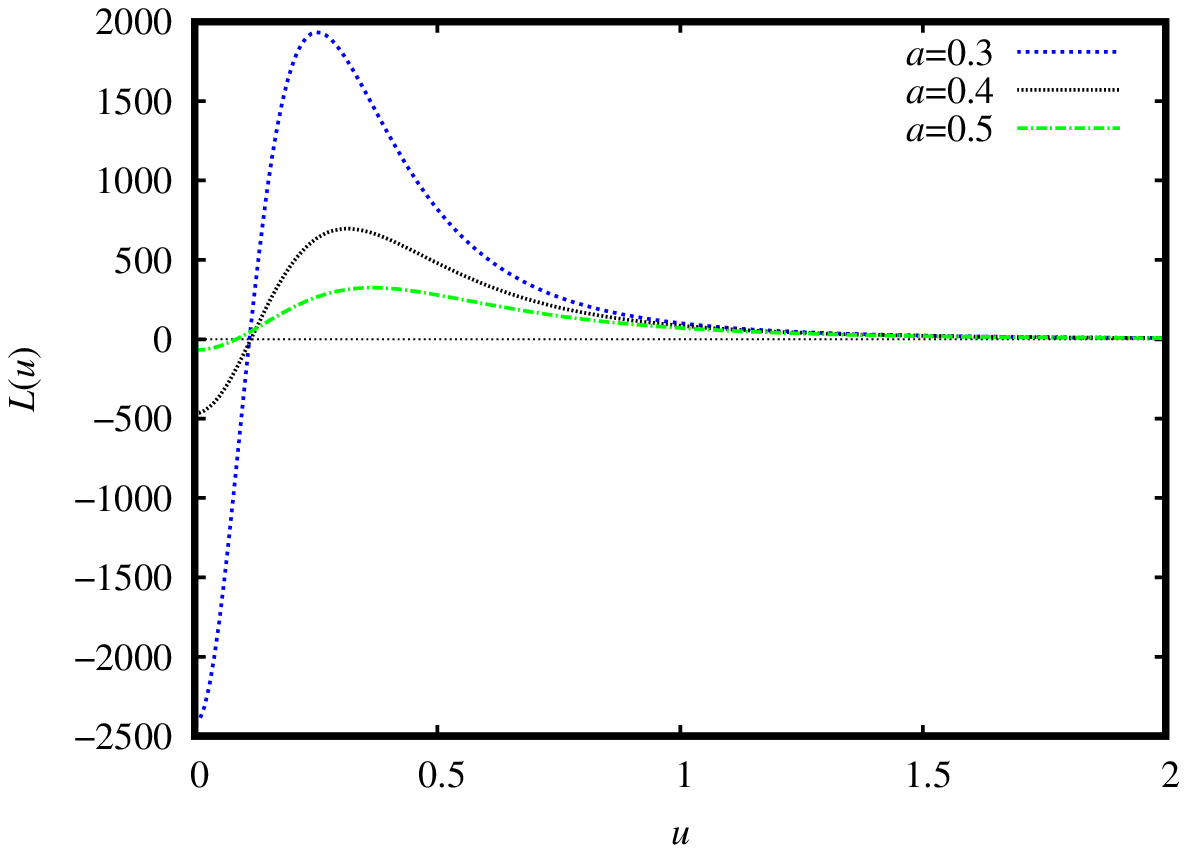}}
\caption{The function $L(F(u))$, Eq.\,\rf{LuInv}, with $k=1$, as a function of the coordinate 
    $u$. In (a) we fixed $a=0.5$ and in (b) we fixed $q=0.5$.}
    \label{fig:LuInverse}
\end{figure}
% --------------------------------------------------
  The function $L(u)$ is plotted in Fig.\,\ref{fig:LuInverse}. The peaks in the electromagnetic  
  Lagrangian become lower as $a$ or $q$ increase. 

\begin{figure}
    \centering
    \subfigure[]{\includegraphics[scale=0.7]{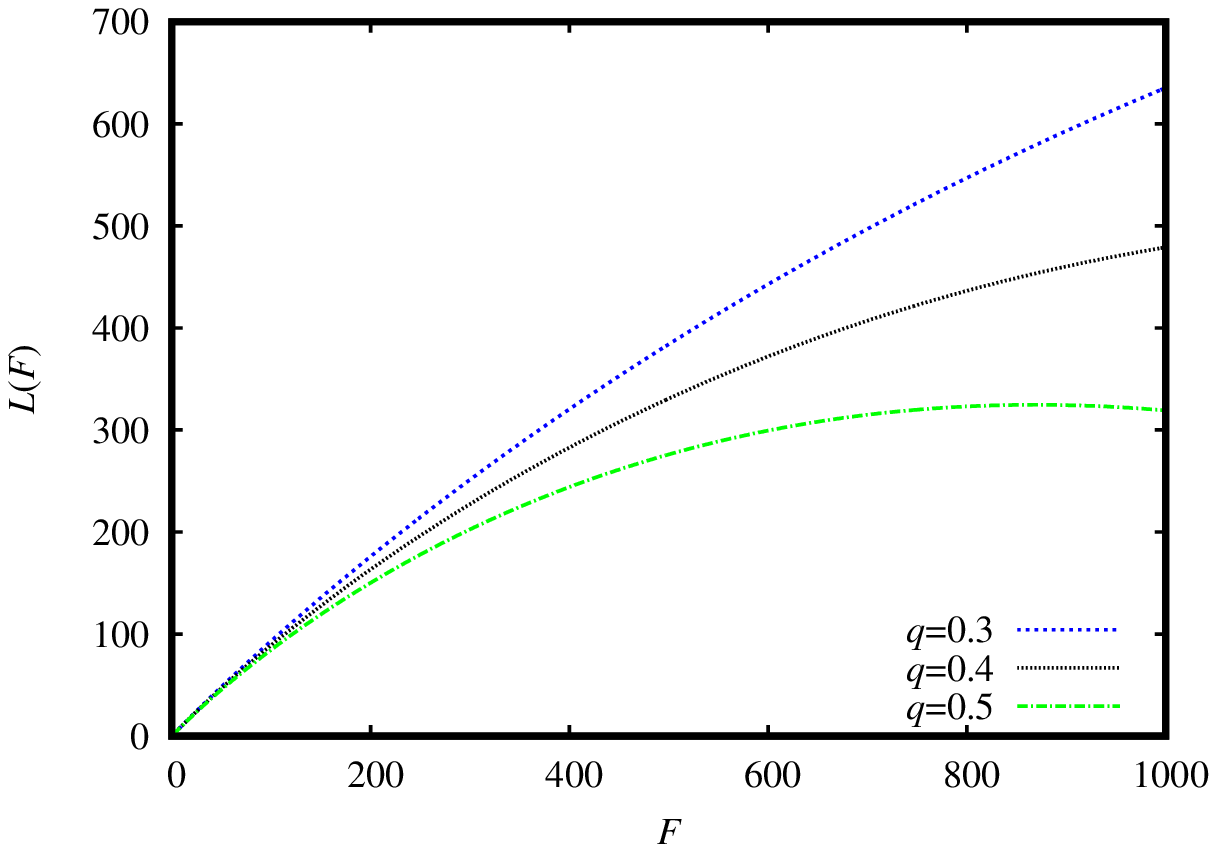}}
    \subfigure[]{\includegraphics[scale=0.7]{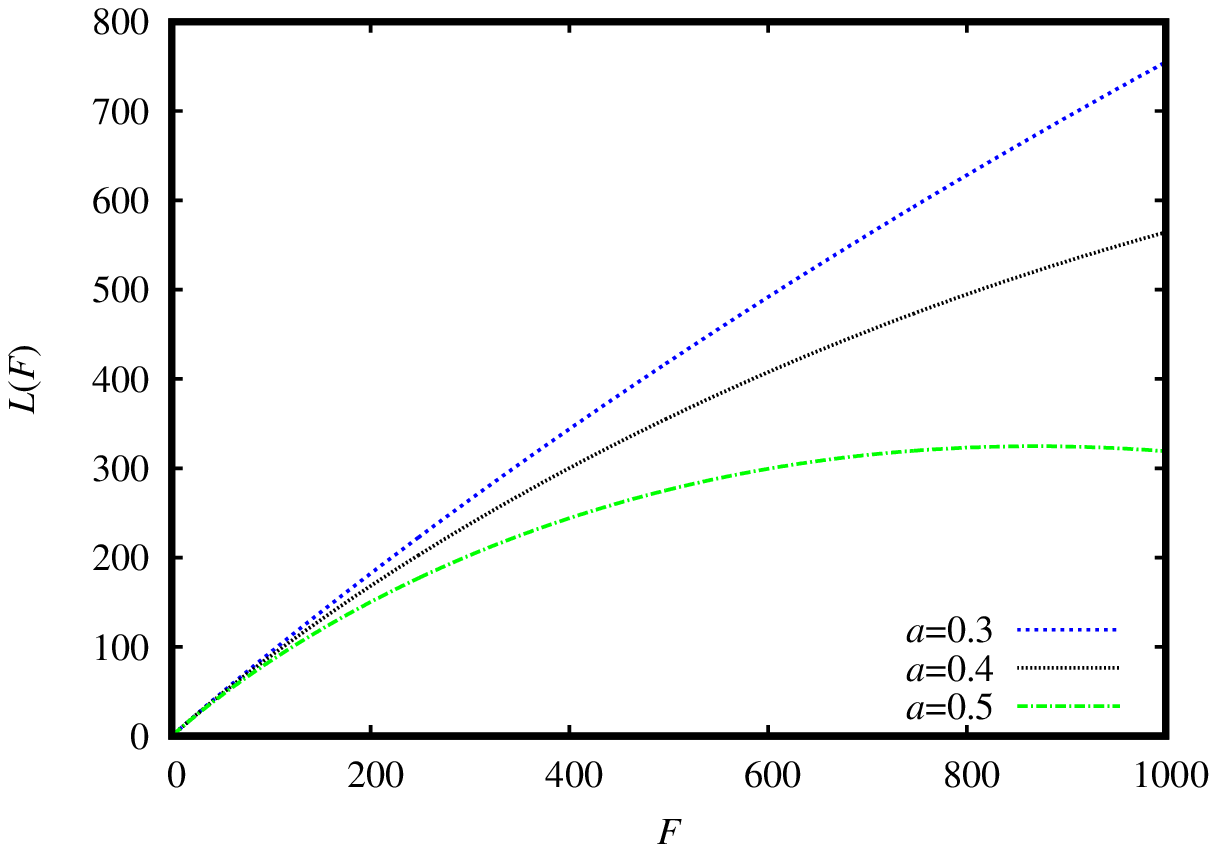}}
    \caption{The function $L(F)$, Eq.\,\rf{LFInv}, with $k=1$. In (a) we fixed $a=0.5$, 
    and in (b) we fixed $q=0.5$.}
    \label{fig:LFInverse}
\end{figure}

\begin{figure}
    \centering
    \subfigure[]{\includegraphics[scale=0.7]{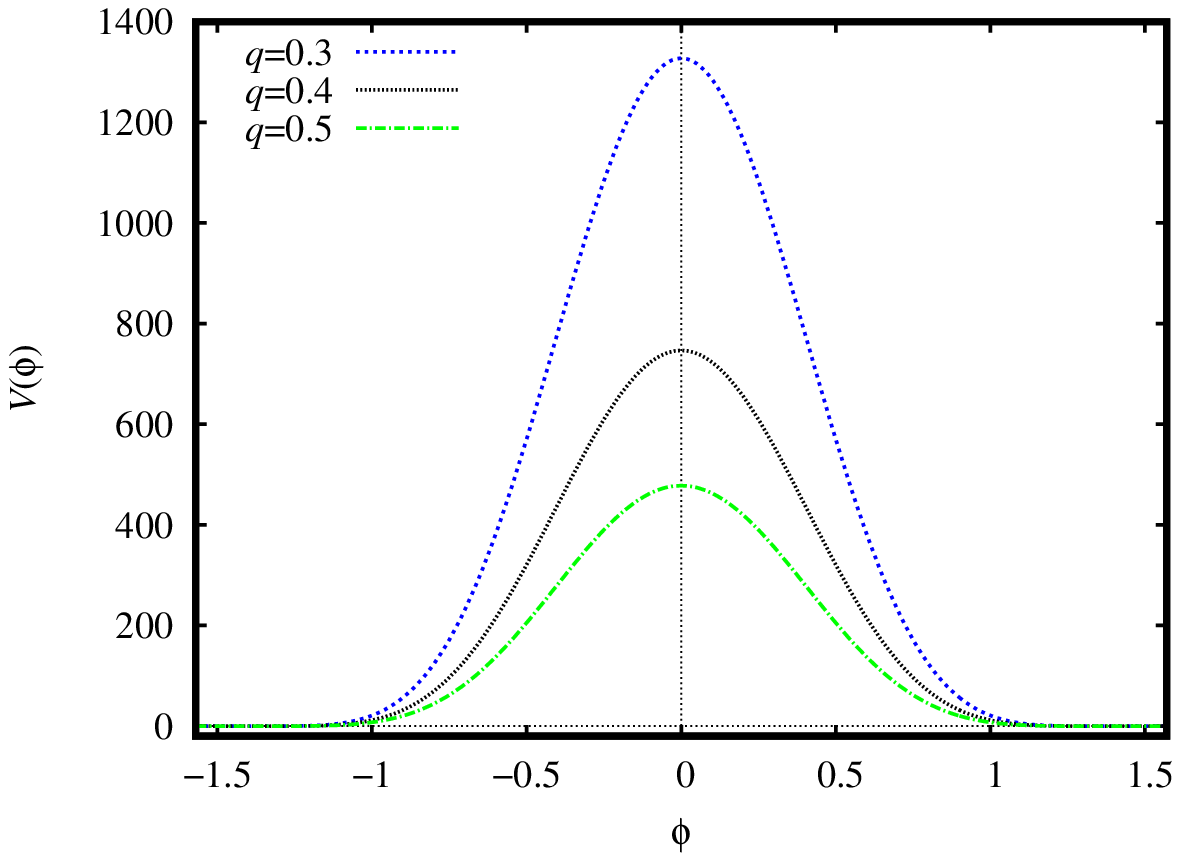}}
    \subfigure[]{\includegraphics[scale=0.7]{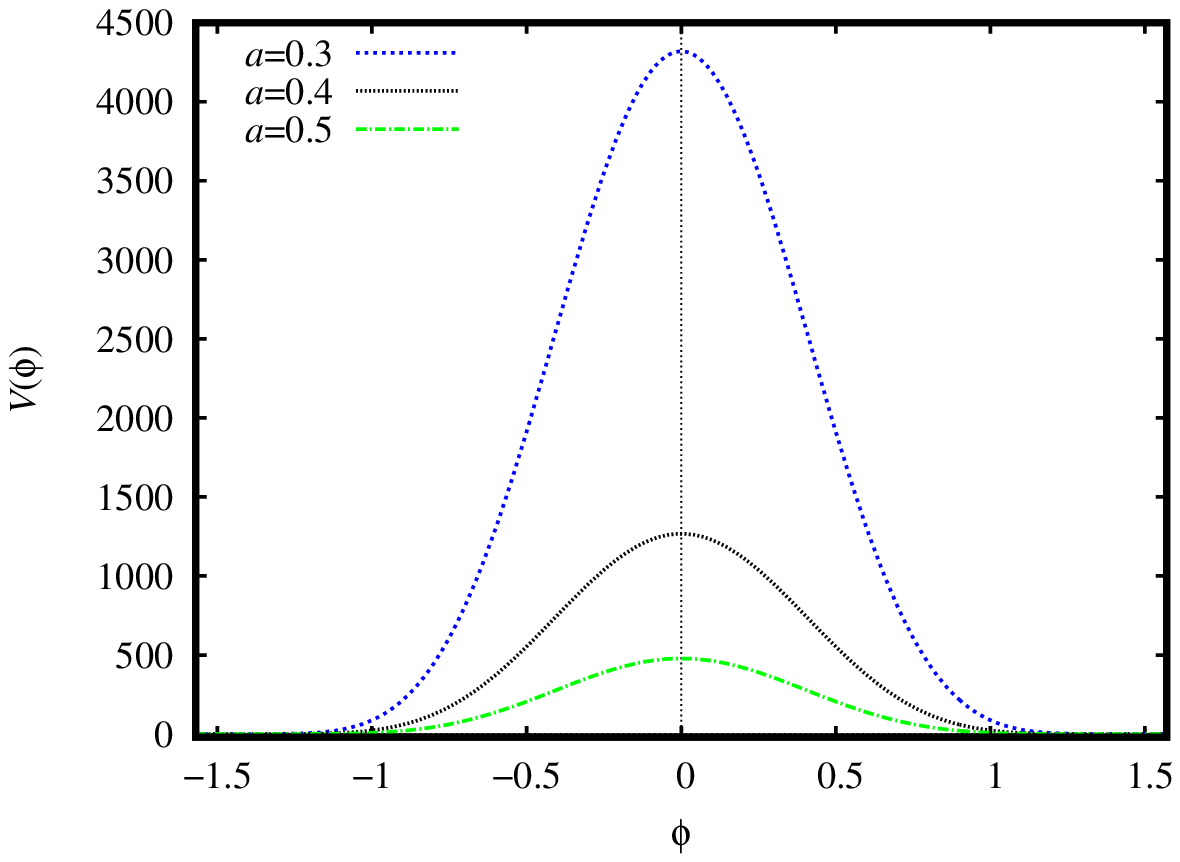}}
\caption{The potential $V(\phi)$, Eq.\,\eqref{VphiInv}, with $k=1$. 
	In (a) we fixed $a=0.5$ and in (b) we fixed $q=0.5$.}
    \label{fig:VphiInverse}
\end{figure}

  The behavior of $L(F)$ and $V(\phi)$ is presented in Figs.\,\ref{fig:LFInverse} and 
  \ref{fig:VphiInverse}. At small $F$, the dominating term in the electromagnetic Lagrangian is 
  $L(F)\approx F$. The first nonlinear term that appears is of the order $F^{5/4}$. The 
  potential looks like a barrier as the scalar field changes. At zero scalar, the potential tends
  to a constant, and the first correction term is proportional to $\phi^2$, 
  simulating a scalar field mass such that $m_\phi^2 = 8(2-a)/(a^2 q^2)$.
% ---------------------------------------------- fig 8
% -------------------------------------------- fig 9
% ----------------------------------------------------

% ===============================
\section{Conclusion}\label{S:conclusion}
% ===============================

  In this paper, we have considered SV-type regularizations for two 
  cylindrically symmetric space-times with Killing horizons: (1) the black string proposed by 
  Lemos \cite{lemos95} (previously analyzed by Lima \textit{et al}. \cite{lima22}) and (2) the so-called 
  inverted black hole [51, 52]. For both families of regularized 
  space-times, we have studied the global causal structures, constructing the appropriate
  Carter-Penrose diagrams, and found possible field sources in the framework of GR, consisting 
  of a phantom minimally coupled scalar field with a nonzero potential and a nonlinear magnetic 
  field. The global regularity of these regularized metrics \rf{string-reg} and \rf{ibh-reg} is 
  verified by finiteness of the Riemann tensor components \rf{Riem} or the Kretschmann
  invariant. 
  
  Next, we proposed a way to find, in the framework of GR, field sources for a class of 
  cylindrically, planarly or toroidally symmetric metrics of the form \rf{ds}, by combining a 
  minimally coupled scalar field $\phi$ with a self-interaction potential $V(\phi)$ and a 
  nonlinear magnetic field described by a certain NED theory with a Lagrangian $L(F)$.
  As both regularized metrics \rf{string-reg} and \rf{ibh-reg} belong to this class, we have
  obtained explicit forms of field sources for both of them.
    
  In particular, the regularized black string solution, which may be called a black-bounce 
  version of the black string proposed by Lemos \cite{lemos95}, has different structures,
  depending on the value of the regularization parameter $a$:  at small $a$ we have a 
  regular black hole with two horizons and a black bounce at $u=0$, and at larger $a$ 
  we obtain a regular black hole with a single extremal horizon at $u=0)$ or a cylindrically 
  symmetric wormhole with a throat at $u=0$. (A similar study was recently carried out
  by Lima et al. \cite{lima23} for the charged version of black strings \cite{lemos96}.)

  In the regularized inverted black hole solution, at $a < 1$ there is a throat at $u=0$ 
  surrounded by two cosmological-type horizons. At $a = 1$, a static region vanishes, and
  we have a regular Bianchi-type I cosmology with an extremal horizon at $u=0$. At $a>1$, 
  we have a regular Bianchi-type I cosmology with a bounce at $u=0$. 
  
  In both regularized solutions, the null energy condition is violated everywhere due to $r''/r>0$,
  and the scalar part of the source is necessarily phantom; still some of the standard energy 
  conditions are respected at least in a certain part of space-time, as can be seen from Fig.\,3.

  The scalar field in both solutions is of kink type and varies between two finite limits, while 
  the potential $V(\phi)$, being everywhere finite, rapidly tends to zero as $|u|\to \infty$,
  although these infinities are of drastically different nature in the two solutions. 
  Meanwhile, the NED sources of the solutions are quite different in nature. In the case of a
  regularized black string, $L(F)$ has no Maxwell weak field limit and has a constant 
  contribution $-6\alpha^2$ (see \eq\rf{LuString}) corresponding to the AdS asymptotic 
  behavior of the metric. Unlike that, in the regularized inverted black hole solution,
  both $V(u)$ and $L(u)$ tend to zero as $|u|\to \infty$, and $L(F)$ has a Maxwell weak 
  field limit.
  
  A common feature of these regular solutions is that the magnetic fields exist without
  their own source (which could be imagined as some distributed current or monopole 
  charge density), their lines of force stretching from one infinity to the other. 
  
  We can conclude that SV-type regularization of known solutions of GR with horizons and 
  singularities leads to a number of geometries of interest, and, in turn, their possible
  field sources shed a certain new light on their properties and can be useful, in particular,
  for studying their stability under various kinds of perturbations.
  
  It is clear that the SV regularization trick can be applied to many other singular metrics,
  including general \cy\ ones, described by \eqn{ds-cy}. Since they contain one more 
  degree of freedom as compared to \rf{ds}, they will require more general field sources
  than those described here, and this can be a subject of a future study.   
  
%==========================================
                                                                                              
\subsection*{Acknowledgements}

M.E.R.  thanks Conselho Nacional de Desenvolvimento Cient\'ifico e Tecnol\'ogico - CNPq, 
Brazil  for partial financial support. 
K.B. acknowledges partial support from the Ministry of Science and Higher Education of
the Russian Federation, Project ``Fundamental properties of elementary particles and 
cosmology No. 0723-2020-0041 and from Project No. FSSF-2023-0003.
%==========================================

%==========================================
\end{document}